\documentclass[]{aa}
\usepackage{graphicx}
\usepackage{epsfig}
\usepackage{subfigure}
\usepackage{amssymb}
\usepackage{natbib}
\bibpunct{(}{)}{;}{a}{}{,}

\newcommand{\FIG}[1]{#1}

\def\mso{\,{\rm M}_\odot}
\def\mst{\,{\rm M}_\star}
 
 \def\rst{\,{\rm R}_\star}

 \def\kms{\, {\rm km}\, {\rm s}^{-1}}

 \def\simle{\mathrel{\hbox{\rlap{\hbox{\lower4pt\hbox{$\sim$}}}\hbox{$<$}}}}
 \def\simgr{\mathrel{\hbox{\rlap{\hbox{\lower4pt\hbox{$\sim$}}}\hbox{$>$}}}}

 \def\msoy{\, \mso~{\rm yr}^{-1}}

\def\vel{\, \mathbf u}
\begin{document}
   \title{Multi-dimensional models of circumstellar shells around evolved massive stars}

   \author{A. J. van Marle
          \inst{1,2}
          \and
          R. Keppens
          \inst{1}
          }

   \offprints{A. J. van Marle}

   \institute{Centre for Plasma Astrophysics, Department of Mathematics, KU Leuven, 
              Celestijnenlaan 200B, B-3001 Heverlee, Belgium \\
              \email{Rony.Keppens@wis.kuleuven.be}    
    \and
         Institute of Astronomy, Department of Physics and Astronomy, KU Leuven, 
              Celestijnenlaan 200D, B-3001 Heverlee, Belgium \\
              \email{AllardJan.vanMarle@ster.kuleuven.be}
}
   \date{Received <date> / Accepted <date>}

\abstract{Massive stars shape their surrounding medium through the force of their stellar winds, which collide with 
the circumstellar medium. Because the characteristics of these stellar winds vary over the course of the evolution of the star, the circumstellar matter becomes a reflection of the stellar evolution and can be used to determine 
the characteristics of the progenitor star. 
In particular, whenever a fast wind phase follows a slow wind phase, the fast wind sweeps up its predecessor in a shell, which is observed as a circumstellar nebula.}
{We make 2-D and 3-D numerical simulations of fast stellar winds sweeping up their slow predecessors to investigate whether numerical models of these shells have to be 3-D, 
or whether 2-D models are sufficient to reproduce the shells correctly.}
{We use the MPI-AMRVAC code, using hydrodynamics with optically thin radiative losses included, to make numerical models of circumstellar shells around massive stars in 2-D and 3-D and compare the results. 
We focus on those situations where a fast Wolf-Rayet star wind sweeps up the slower wind emitted by its predecessor, being either a red supergiant or a luminous blue variable.}
{As the fast Wolf-Rayet wind expands, it creates a dense shell of swept up material that expands outward, driven by the high pressure of the shocked Wolf-Rayet wind. These shells are subject to a fair variety of hydrodynamic-radiative instabilities. 
If the Wolf-Rayet wind is expanding into the wind of a luminous blue variable phase, the instabilities 
will tend to form a fairly small-scale, regular filamentary lattice with thin filaments connecting knotty features. If the Wolf-Rayet wind is sweeping up a red supergiant wind, the instabilities will form larger interconnected structures with less regularity.
The numerical resolution must be high enough to resolve the compressed, swept-up shell and the evolving instabilities, which otherwise may not even form.} 
{Our results show that 3-D models, when translated to observed morphologies, give  realistic results that can be compared directly to observations. The 3-D structure of the nebula will help to distinguish different progenitor scenarios.}

  \titlerunning{Circumstellar shells around evolved massive stars}
  \authorrunning{van Marle \& Keppens}

   \keywords{Hydrodynamics -- Instabilities --
                Methods: numerical -- 
                Stars: circumstellar matter --
                Stars: massive --
                Stars: winds: outflows
               }

  \maketitle

%

\section{Introduction}
As a massive star evolves, it loses a large fraction of its mass in the form of stellar wind. 
The parameters of the wind change during the course of stellar evolution \citep[e.g.][]{deJageretal:1988}, 
leading to a series of interactions in the circumstellar medium (CSM). 
Whenever the wind parameters make a transition from a slow to a fast phase, the fast wind will collide with its 
slower predecessor, sweeping the older material up in a moving shell.
These shells, known as circumstellar nebulae, can be observed directly. 
This will be either in emission, if the shells are ionized by UV radiation from the central star, or in absorption, as 
Doppler shifted absorption features in the spectrum of a separate source such as the central star. 
Since the shells are a direct result of the changes in wind parameters, which in turn are caused by the evolution of the star, 
the morphology of the CSM becomes a unique tool to determine the evolutionary path of the central star. 
A typical example of such an interaction occurs when a massive star ($\gtrsim30\,$M$_\odot$) makes the transition from giant star to Wolf-Rayet (WR) star. 

As a star reaches the end of its giant phase, it makes a rapid transition from the cool to the hot part of the Hertzsprung-Russell diagram to become a WR star \citep[e.g.][]{Langeretal:1994}.
This transition causes a radical change in the wind parameters \citep[][and references therein]{Lamerscassinelli:1999}. 
During the giant phase, the wind is relatively slow ($10-200\,\kms$), due to the low escape velocity. 
The WR star, which has a much smaller radius, has a high escape velocity, leading to a fast wind ($\sim\,2000\,\kms$). 
As this fast wind encounters its slow predecessor, it creates a thin, dense shell, which can be observed as a circumstellar nebula. 
Such nebulae, known as WR ring nebulae, have been found around many WR stars \citep{MillerChu:1993}.

This process is similar to formation of a planetary nebula, which occurs when a low mass star makes the transition from Asymptotic Giant Branch (AGB) to post-AGB star \citep{Kwoketal:1978}. 
However, due to the much higher kinetic energy of the WR wind as compared to the wind of a post-AGB star, the shells of WR nebulae tend to be driven by an energy conserving, 
rather than a momentum conserving interaction, leading to higher expansion velocities \citep[chapter 12][]{Kwok:2000}. 
Also, WR stars produce very large amounts of high energy photons, allowing them to fully ionize their nebulae \citep{vanMarleetal:2005,vanMarleetal:2007,ToalaArthur:2011},
 rather than the partial ionization expected from post-AGB stars \citep{Garcia-Seguraetal:1999}. 

These circumstellar shells, which are pushed by the thermal pressure of the shocked WR wind, are subject to linear thin-shell instabilities \citep{Vishniac:1983}. 
Since the material in the shell is also much denser than the shocked wind that pushes it outward, it can also form Rayleigh-Taylor (RT) instabilities. 
As a result, the structure of these shells can become very complicated as has indeed been observed in, 
for example, \object{RCW~58}, \object{RCW~104}, \object{NGC~3199} and \object{NGC~6888} \citep[E.g.][]{Chuetal:1983,Goudisetal:1988,Smithetal:1988,DysonGhanbari:1989,Gruendletal:2000}. 
Therefore, it becomes necessary to simulate the interaction in more than one dimension.  

As a massive star reaches the end of its evolution it will die, 
usually in a spectacular fashion, creating a supernova and/or gamma-ray burst (GRB). 
These violent events produce fast moving shocks that expand in the CSM. 
The presence of a circumstellar shell can be observed in several ways while studying a supernova or GRB. 
Like the central star, the supernova or GRB illuminates the CSM with high energy radiation. 
This ionizes the shell, which then in turn becomes visible as electrons and ions recombine. 
This effect is present in the double ring nebula around SN~1987A \citep{Burrowsetal:1995}, which may well be the key to understanding the exact nature of its progenitor 
\citep[e.g.][]{MorrisPodsiadlowski:2005,Chitaetal:2008}. 
The presence of a circumstellar shell can also be observed as a distinct, blue-shifted absorption feature in the spectrum of some supernovae like \object{SN~1998S} \citep{Bowenetal:2000,Fassiaetal:2001}, 
as its velocity is different from the stellar winds.
For GRBs this is more difficult to observe, as the high energy photons from the GRB and its afterglow will ionize the surrounding gas to a very high degree \citep{Prochaskaetal:2007}, 
though the presence of dust in the CSM may compensate for this to some extent \citep{Robinsonetal:2010}. 
Alternatively, it may be possible to find absorption lines at higher energy levels, if the GRB itself is red-shifted sufficiently to bring them within the IR-optical part of the spectrum \citep{Prochaskaetal:2008}.

If an expanding supernova comes in direct contact with a circumstellar shell, the interaction can 
completely change the circumstellar environment, depending on the ratio of mass between the supernova and the circumstellar nebula 
as demonstrated in simulations by, for example, \citet{TenorioTagleetal:1990,TenorioTagleetal:1991,Rozyczkaetal:1993,ChevalierDwarkadas:1995,Dwarkadas:2005,Dwarkadas:2007} and 
\citet{vanVeelenetal:2009}. 
This was confirmed observationally by \citet{McCray:2005}. 
When the supernova hits the shell, the increased density at the shockfront will create extra radiation causing a rise in the supernova lightcurve as demonstrated by \citet{vanMarleetal:2010}. 
The collision between a supernova and a circumstellar shell was predicted in the case of \object{SN1987A} by \citet{ChevalierLiang:1989} and \citet{LuoMcCray:1991}. 
Observations in  X-ray \citep{Parketal:2005}, optical \citep{McCray:2005} and infra-rad \citep{Boucheetal:2006} show brightening due to interaction between 
the supernova and the ionized region interior to the circumstellar shell, which is denser than the wind. 
As the supernova remnant expands further we can expect to observe its interaction with the shell itself. 
Whether such an interaction can be observed for a GRB is more doubtful, though the transition from a $1/r^2$ (wind-like) to a constant density medium can be visible, 
depending on the opening angle of the GRB jet \citep[e.g.][]{vanEertenetal:2010}. 

In order to infer the stellar evolution from the observable characteristics of the circumstellar environment, we need to make detailed numerical models of the wind interactions. 
This has been done previously by \citet{GarciaSeguraetal:1996b,vanMarleetal:2005,Freyeretal:2006,Dwarkadas:2007} for the interaction between a WR wind and a red supergiant (RSG) wind 
and by \citet{GarciaSeguraetal:1996a,vanMarleetal:2007,Freyeretal:2003,ToalaArthur:2011} for the interaction between a WR wind and a luminous blue variable (LBV) wind. 
Many such simulations were limited to 2-D models, as
was the large grid of models computed by \citet{Eldridgeetal:2006}. 
\citet{Chitaetal:2008} and \citet{vanMarleetal:2008} added the effect of rotation and
recently, \citet{ToalaArthur:2011} improved on the earlier 2-D models by introducing new physics in the form of radiative transfer and thermal conduction. 
Although the effect of thermal conduction proved insignificant, \citet{ToalaArthur:2011} showed that radiation
from the star can play an important part in shaping the circumstellar nebula by ionizing the shell.
These yield a lot of useful insight in the evolution of the CSM, but, since they are all limited to 2-D, they cannot fully describe what is fundamentally a 3-D process. 
A single 3-D model was shown in \citet{vanMarleetal:2011a}. 
Here we present several simulations at varying resolution to investigate the difference between the results of 2-D and 3-D simulations to determine if and when 3-D models are necessary. 

The use of the {\tt{MPI-AMRVAC}} code \citep[][and references therein]{Melianietal:2007,vanderHolstetal:2008,Keppensetal:2012}, 
which can solve the hydrodynamics equations for relativistic as well as classical hydrodynamics, 
also helps us prepare for future work, in which the models of the CSM can be combined with simulations of supernovae and GRBs. 
As input models we use two stellar wind interactions, both of which involve a fast wind sweeping up its slower predecessor. 
In the first case, a WR wind sweeps up its RSG predecessor. 
In the second, the WR wind sweeps up an earlier LBV wind. 
For our input models we use wind parameters based on recent work by \citet{Bouretetal:2005,Vinkdekoter:2005,Mokiemetal:2007}. 
These reflect the latest developments in our understanding of the winds of massive stars. 
Keeping the physical parameters of each model constant, we make both 2-D and 3-D simulations and vary the resolution of each model. 
By comparing the end results we determine the effect of a change in resolution and whether 3-D models make a useful contribution over the existing 2-D simulations. 

In the following sections, we describe the numerical method we use to simulate the interactions (Section~\ref{sec-meansmethods}) and  present the results (Section~\ref{sec-results}). 
We discuss the instabilities that appear in our models in Section~\ref{sec-instabilities}. 
Using the 3-D models, we show how they would appear to an observer in Section~\ref{sec-observation}.
Finally, in Section~\ref{sec-discussion} we will analyse and discuss the results and in Section~\ref{sec-conclusions} present our conclusions.

  \begin{table*}
      \caption[]{Simulation parameters. Simulations labelled with A represent the WR-RSG interaction, simulations labelled with B represent the WR-LBV interaction.}
         \label{tab:parameters}
\centering
\begin{tabular}{lccccccc}
\hline\hline
Simulation  &  $\dot{M}_{\rm WR}$       & $\dot{M}_{\rm giant}$  & $v_{\rm WR}$ & $v_{\rm giant}$ & Domain size & Effective grid    \\
            &  [M$_\odot /yr]$          & [M$_\odot /yr]$        & [km/s]            & [km/s] &[pc$\,\times\,^{\rm o}\,\times\,^{\rm o}]$    &    \\
\hline
\\
A1         &  $1\times 10^{-5}$ & $1\times 10^{-4}$ &  2000      & 10  & $5\times45$          &  $800\times128$            \\
B1         &  $1\times 10^{-5}$ & $5\times 10^{-4}$ &  2000      & 200 & $5\times45$          &  $800\times128$            \\
A2         &  $1\times 10^{-5}$ & $1\times 10^{-4}$ &  2000      & 10  & $5\times45$          &  $1600\times256$           \\
B2         &  $1\times 10^{-5}$ & $5\times 10^{-4}$ &  2000      & 200 & $5\times45$          &   $1600\times256$          \\
A3         &  $1\times 10^{-5}$ & $1\times 10^{-4}$ &  2000      & 10   & $5\times45\times45$  &   $800\times128\times128$  \\
B3         &  $1\times 10^{-5}$ & $5\times 10^{-4}$ &  2000      & 200   & $5\times45\times45$  &   $800\times128\times128$ \\
A4         &  $1\times 10^{-5}$ & $1\times 10^{-4}$ &  2000      & 10  & $5\times45\times45$  &   $1600\times256\times256$  \\
B4         &  $1\times 10^{-5}$ & $5\times 10^{-4}$ &  2000      & 200  & $5\times45\times45$  &   $1600\times256\times256$ \\
\hline
\end{tabular}
 \end{table*}

\begin{figure}
\FIG{
\centering
\includegraphics[width=0.70\columnwidth]{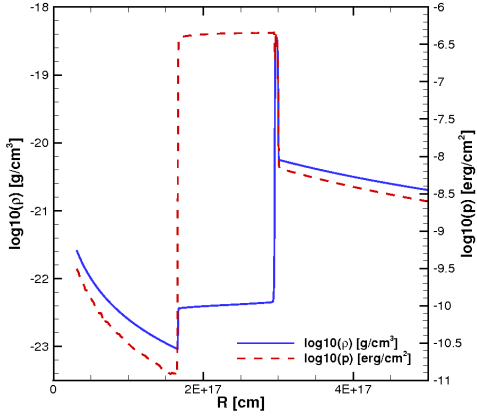}
}
\caption{Density and thermal pressure in the CSM of a WR 100~years after the end of the RSG phase. 
This is the starting point for the simulations labelled with A.}
\label{fig:WR_RSG_1-D}
\end{figure}

\begin{figure}
\FIG{
\centering
\includegraphics[width=0.70\columnwidth]{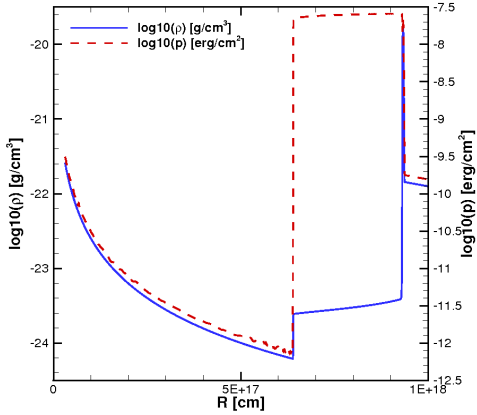}
}
\caption{Similar to Fig.~\ref{fig:WR_RSG_1-D}, but for the transition from LBV to WR. 
This is the starting point for the simulations labelled B.}
\label{fig:WR_LBV_1-D}
\end{figure}

\section{Numerical setup}
\label{sec-meansmethods}
\subsection{Governing equations}
\label{sec-hydroeqns}
We simulate hydrodynamical interactions in the CSM by solving the 
conservation equations for mass:  
\begin{equation}
\frac{\partial \rho}{\partial t} ~+~ \nabla \cdot (\rho \vel) ~=~ 0,
\end{equation}
momentum:  
\begin{equation}
\frac{\partial}{\partial t}( \rho \vel) ~+~\nabla \cdot (\rho \vel \vel) ~=~ -\nabla p ,
\end{equation}
and energy: 
\begin{equation}
\frac{\partial e}{\partial t} ~+~\nabla \cdot (e {\vel} ) ~+~ \nabla \cdot (p\vel) ~=~ -\biggl(\frac{\rho}{m_h}\biggr)^2 \Lambda(T), 
\label{eq:energy}
\end{equation}
with $\rho$ the density, $\vel$ the velocity, $p$ the thermal pressure, $e$ the total energy density and 
$m_h$ the hydrogen mass.
The last equation includes the effect of radiative cooling, which depends on local density, temperature 
and metallicity. 
The exact calculation of the radiative cooling would actually be:
\begin{equation}
\frac{\partial e}{\partial t}~=~-n_e n_i \Lambda(T) ,
\end{equation}
with $n_e$ the free electron particle density and $n_i$ the ion particle density. 
Using this equation would require detailed knowledge of the local composition and ionization state of the gas. 
For a first approximation we have chosen to assume that the gas is fully ionized, which seems acceptable in the light of results by \citet{ToalaArthur:2011}, and that hydrogen is the only 
relevant contributor to the particle density, which simplifies the cooling equation to the form shown in Eq.\,(\ref{eq:energy}).
The cooling function $\Lambda(T)$ is a temperature dependent, which has to be taken from a pre-calculated table, taking into 
account the composition of the gas as well as the ionization states at different temperatures. 
We have chosen to use a table for solar metallicity from \citet{MellemaLundqvist:2002}, which is based on a numerical calculation of the energy loss of ionized gas 
as a function of temperature. 
Using a solar-metallicity based cooling curve is justified in that most of the cooling takes place in the swept-up shell (due to its high density) of RSG or LBV wind material, 
which can be expected to have the same metallicity as the progenitor star had at birth. 

We don't take into account the effect of thermal conduction, but the results from \citet{ToalaArthur:2011} show that this is not a significant factor. 

\subsection{Basic stellar wind parameters}
As test cases we simulate the interaction between a fast WR type star wind and the wind remnant from the earlier RSG phase, and the interaction between a WR wind and an earlier LBV wind. 
The RSG wind is slow and very dense ($V=10\kms$, $\dot{M}=10^{-4}\msoy$), the LBV wind, while faster ($V=200\kms$), has a similar high massloss rate ($\dot{M}=5\times10^{-4}\msoy$). 
The WR wind is much faster ($2\,000\kms$), and has a lower massloss rate ($\dot{M}=10^{-5}\msoy$). 
These particular numbers are based on previous simulations \citep{GarciaSeguraetal:1996a,GarciaSeguraetal:1996b, vanMarleetal:2005,vanMarleetal:2007}, but use a lower massloss rate for the LBV and WR winds as 
argued by \citet{Bouretetal:2005,Vinkdekoter:2005,Mokiemetal:2007}. 
For the full parameters of our simulations, see Table~\ref{tab:parameters}.  

Since the WR star would be a very powerful source of high energy radiation, we assume that the CSM is fully ionized, leading to a minimum temperature of 10,000\,K throughout the entire computational domain. 
This simplification can be justified by analytical approximation, such as done by \citep{McKeeetal:1984} and \citet[][p.\,70]{Dysonwilliams:1997}, which shows that a massive star can ionize the surrounding medium out to several parsec, 
by numerical results from \citet{Freyeretal:2003,Freyeretal:2006,vanMarleetal:2005,vanMarleetal:2007}, which incorporated a simplified form of photo-ionization and recent work by \citet{ToalaArthur:2011}, which included radiative transfer. 
Observations of \object{NGC~6888}by \citet{Mooreetal:2000} also show that the shell is photo-ionized by radiation from the star. 

\subsection{Code}
For our simulations we use the {\tt{MPI-AMRVAC}} code \citep[][and references therein]{Melianietal:2007,vanderHolstetal:2008,Keppensetal:2012}.
This is a state of the art software package that solves the conservation equations on 
grids where the local resolution can be changed through adaptive mesh refinement (AMR). 
Together with the extensive use of MPI, this allows us to efficiently compute large, multi-D simulations. 
Both the MPI and AMR are absolute necessities for running these simulation. 
E.g. the high-resolution 3-D simulations typically take approximately 72-hours, running on 128 processors
on either the VIC3 supercomputer at the Flemish High Performance Computer Centre or the CINECA SP6 in Bologna, Italy. 
Different physics modules let us adapt the code to a specific problem which can include magnetic fields 
and relativistic effects. 
We have recently added a separate module to include the effect of optically thin radiative cooling \citep{vanMarleKeppens:2011}, 
this module uses the exact integration method by \citet{Townsend:2009}, which improves calculation speed and 
numerical stability. 

\subsection{Grid}
We start all simulations in 1-D, using a spherically symmetric grid. This allows the shock to form without numerical instabilities. 
Our initial 1-D grid has a maximum radius of 5~pc and a basic grid of 400~points. 
We allow 7 additional levels of refinement, which gives us an effective resolution of 51\,200~points. 
We initialize the simulation by filling the entire grid with wind material according to the parameters of the slow (RSG/LBV) wind. 
At the inner  radial boundary we allow  gas to flow in according to the parameters of the fast WR wind. 
The transition from slow wind to fast wind is assumed to be instantaneous, rather than gradual. 
This is a reasonable assumption, since the transition happens on the typical free-fall timescale of a star,
\begin{equation}
 t_{\rm ff}~\propto~\frac{\rst}{v_{\rm esc}}~=~\sqrt{\frac{\rst3}{2G \mst}},
\end{equation}
with $v_{\rm esc}$, $\rst$ the stellar radius, $\mst$ the stellar mass, and $G$ Newton's gravitational constant.
Even for supergiants, this works out to about $10^5-10^7$\,s, which is much shorter than the timescales on which the nebulae develop (typically thousands of years).

\citet{Dwarkadas:2005,Dwarkadas:2007} found that starting a simulation in 1-D influences the final result because turbulence in the shocked wind region 
can influence the shape of the wind termination shock. 
This changes the morphology of a RSG shell formed against the shock and does not occur in a 1-D simulation. 
This is not a problem in our case, because we do not simulate the formation of such a RSG shell, which occurs outside the physical domain of our models. 
Also, we only simulate the first 100 years in 1-D before we change to 2-D or 3-D. 
As the 2-D or 3-D simulation starts, we seed the domain with random density variations.  
Thereafter, any side-effects of turbulent shocked wind are taken into account.

After about 100~years, once the shock is formed and starts to move outward, we map the 1-D result onto a 2-D grid. 
This approach helps us to avoid numerical problems \citep[such as those described by][]{Quirk:1994}. 
Our grids are centred on the equator with a latitudinal opening angle of $45^{\rm o}$ ($22.5^{\rm o}$ above and below the equatorial plane). 
The 3-D simulations follow the same pattern with a $45^{\rm o}$ latitudinal opening angle. 
For these simulations the maximum radius is again 5~pc. 
The basic radial grid is 400~points, with either 1 or 2 additional levels of refinement. 
The basic angular resolution is 64 gridpoints. 
(For the resulting effective grid resolutions see Table~\ref{tab:parameters}).

For our 1-D simulations we use the Total Variation Diminishing, Lax-Friedrich (TVDLF) scheme  \citep{TothOdstrcil:1996}, combined with `minmod' limiting. 
For the 2-D and 3-D simulations, we replace the `minmod' limiter with the more accurate `van Leer' limiter method \citep{vanLeer:1974} to compensate for the lower resolution. 
In order to facilitate the formation of instabilities, we randomly seed both the slow and fast wind with a one-percent density variation throughout the entire domain.

\begin{figure*}
\FIG{
 \centering
\mbox{
\subfigure
{\includegraphics[width=0.5\textwidth]{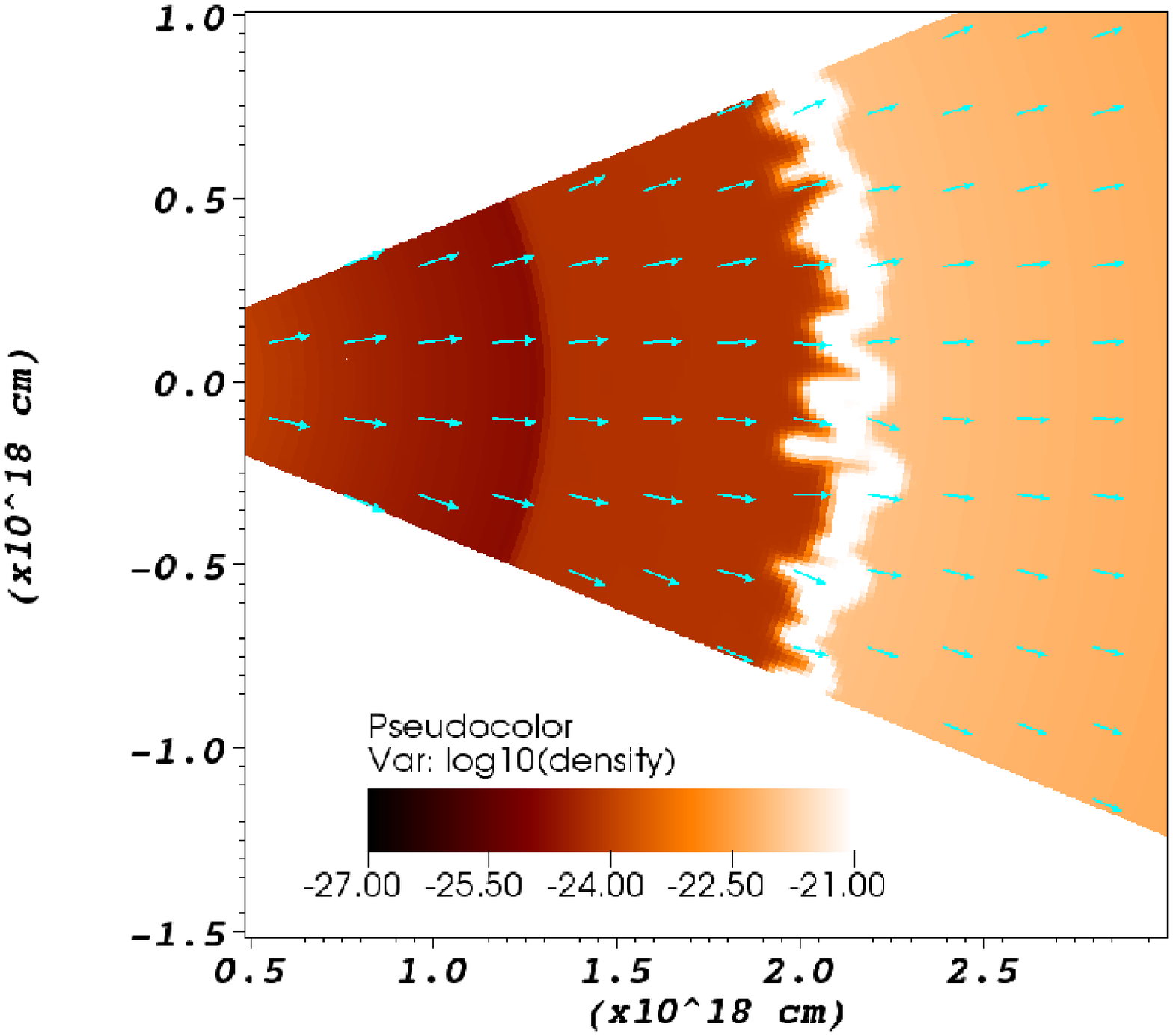}}
\subfigure
{\includegraphics[width=0.5\textwidth]{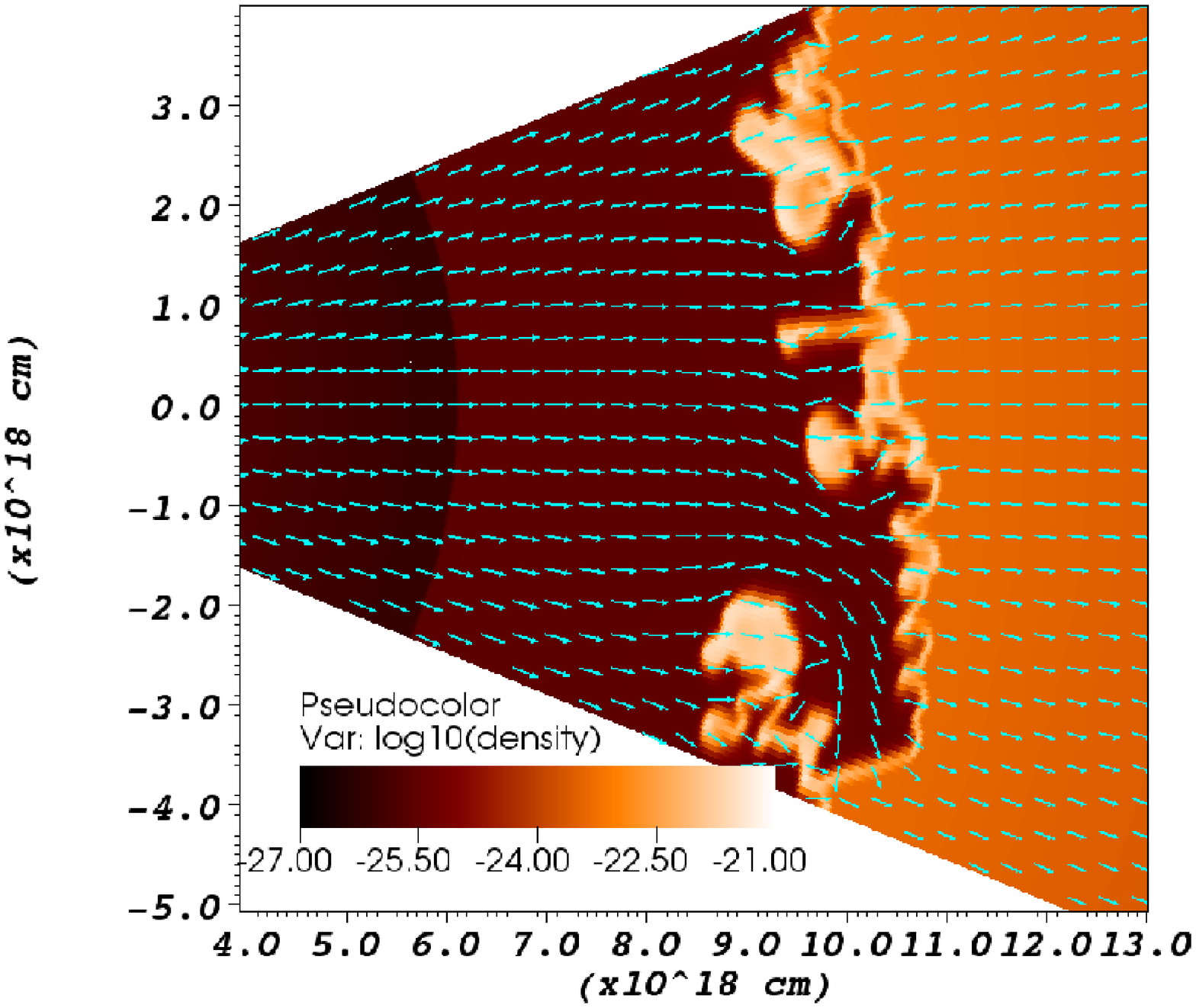}}}
}
\caption{The WR-RSG interaction in low resolution, showing density in cgs of the CSM 
and the velocity field for simulation A1 after 7\,920 (left panel) and 39\,200 years (right panel). 
At first, the shell clearly shows the development of linear Vishniac instabilities. 
As the shell moves outward, RT instabilities appear as well, leading to large scale disruption of spherical symmetry. 
In the right panel, both Vishniac and RT instabilities remain visible.}
 \label{fig:WR_RSG_2-D_low}
\end{figure*}

\begin{figure*}
\FIG{
 \centering
\mbox{
\subfigure
{\includegraphics[width=0.5\textwidth]{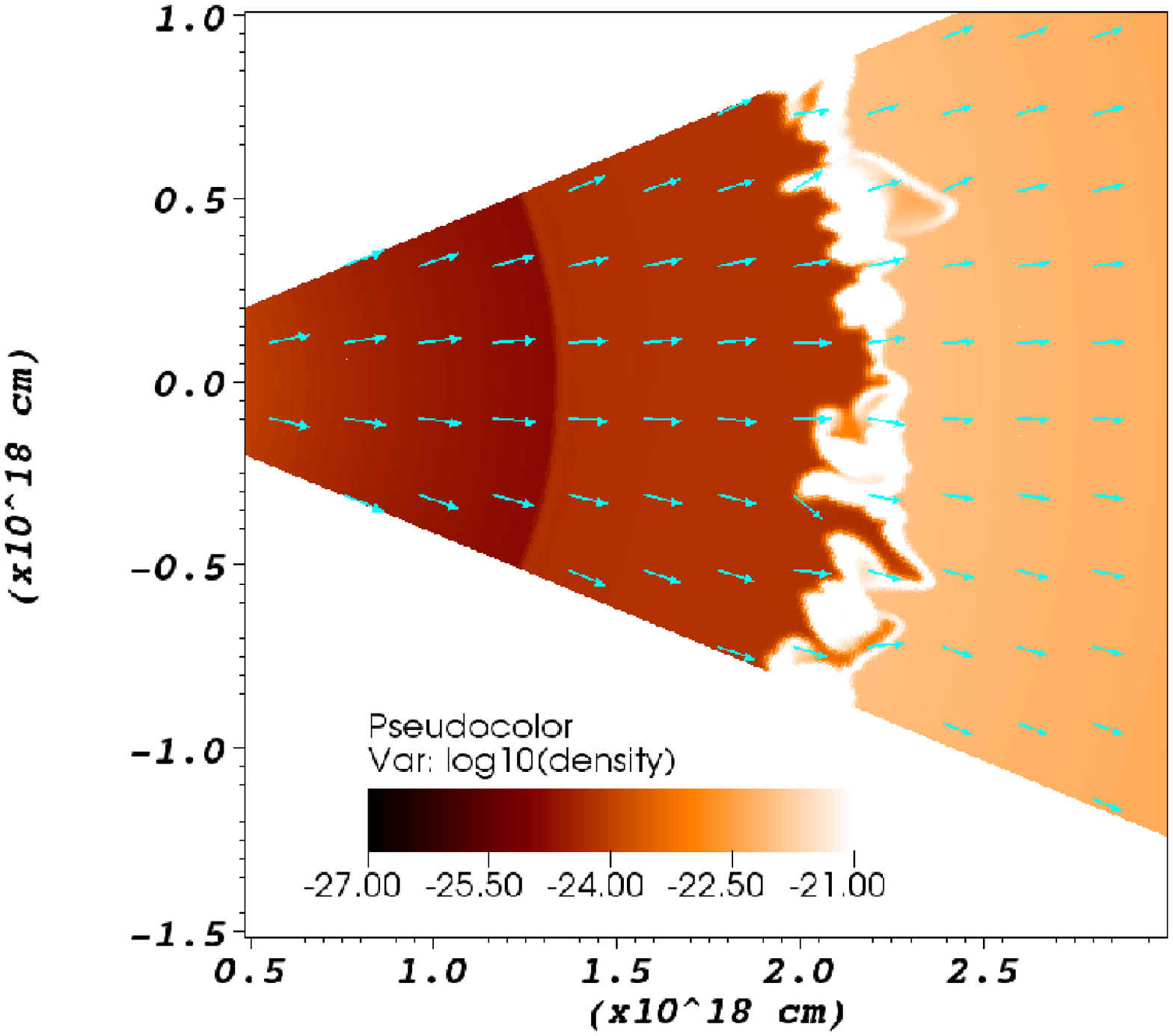}}
\subfigure
{\includegraphics[width=0.5\textwidth]{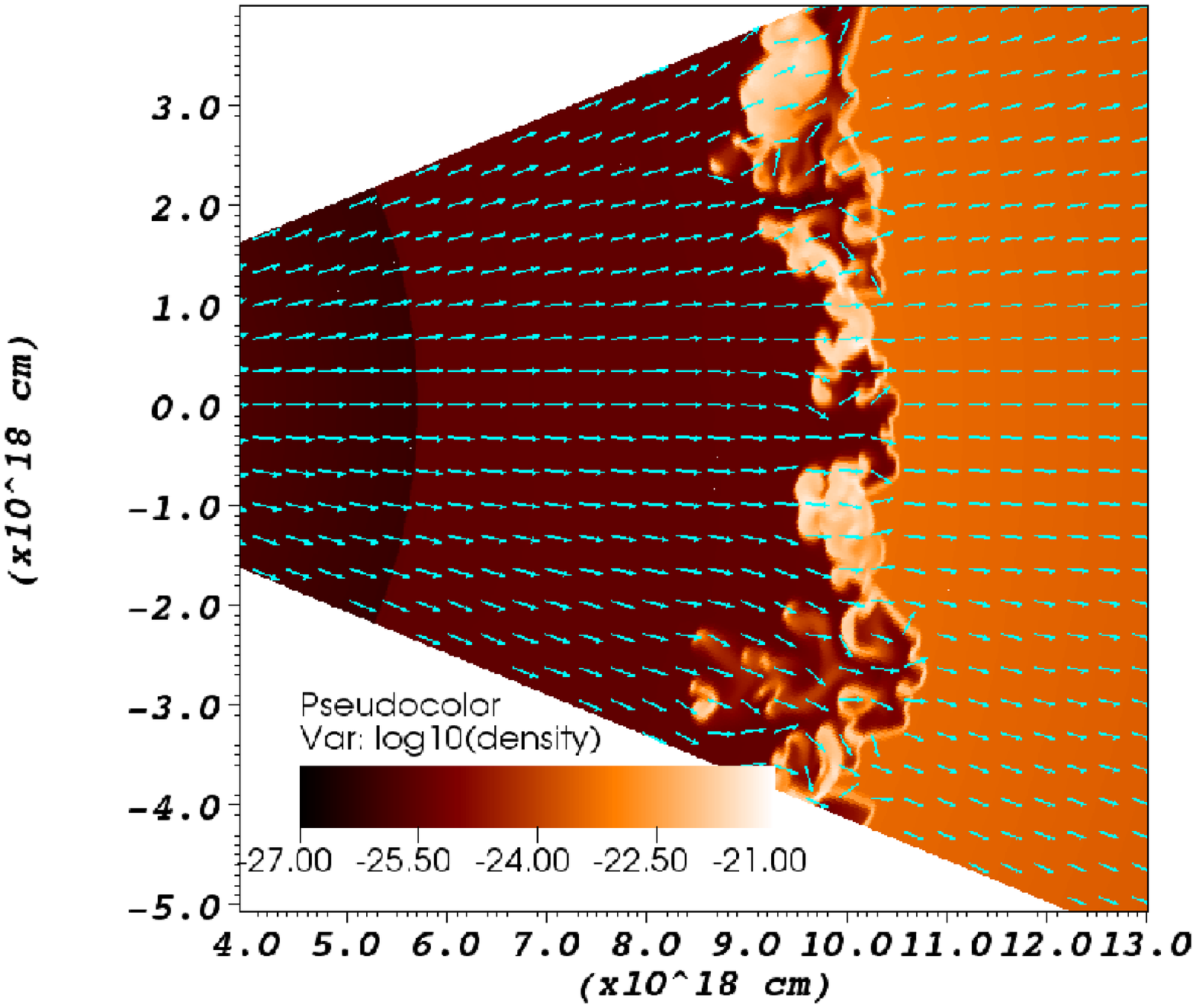}}}
}
\caption{Similar to Fig.~\ref{fig:WR_RSG_2-D_low}, but for simulation A2, which has twice the resolution. Initially, the difference with the low resolution run is small. 
The RT instabilities in the right panel are smaller than their counterparts in Fig.~\ref{fig:WR_RSG_2-D_low} and show more structure. 
Parts of the RT fingers are breaking off and will eventually dissolve in the shocked WR wind bubble that is driving the shell.}
 \label{fig:WR_RSG_2-D_high}
\end{figure*}

\begin{figure*}
\FIG{
 \centering
\mbox{
\subfigure
{\includegraphics[width=0.5\textwidth]{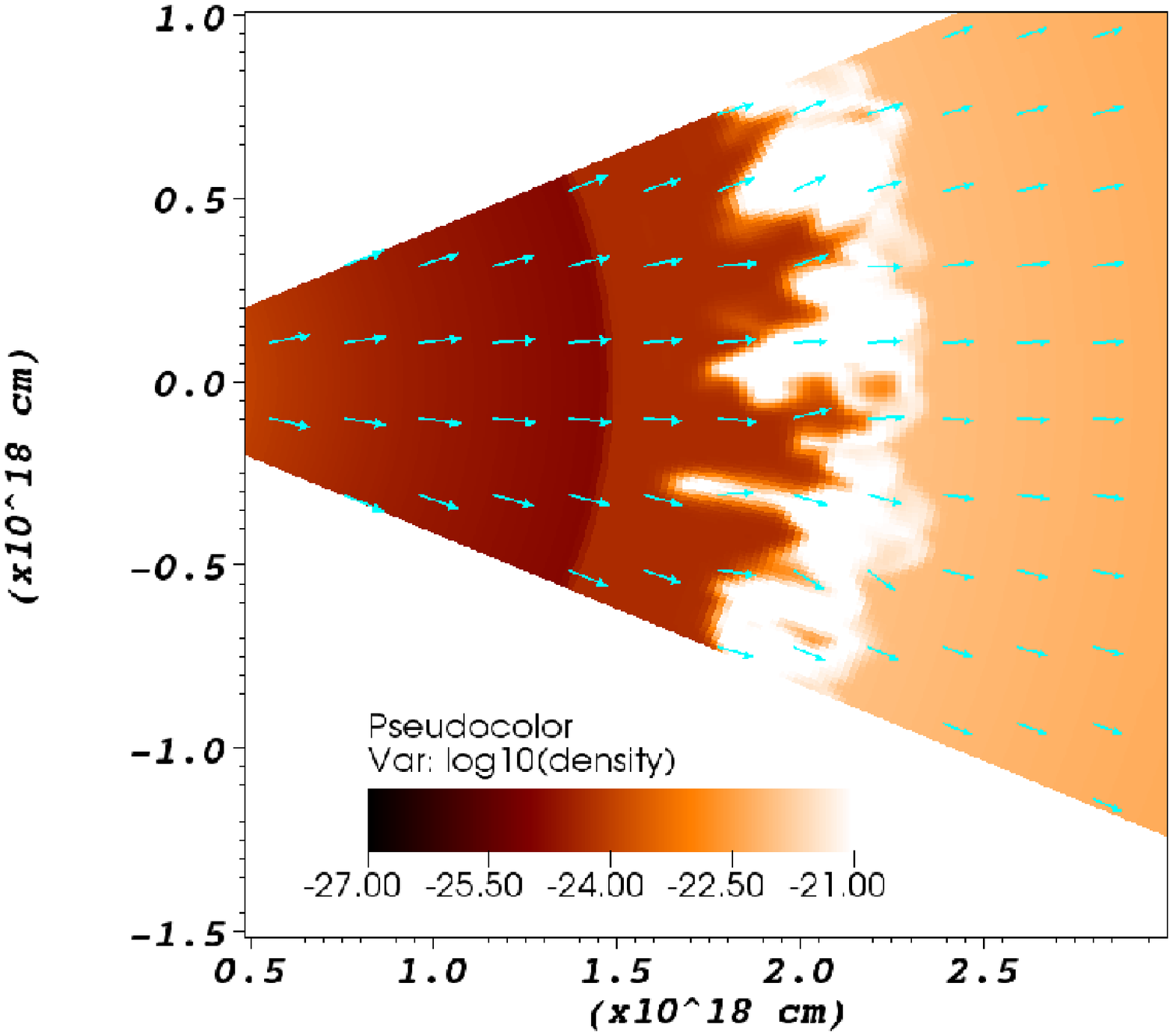}}
\subfigure
{\includegraphics[width=0.5\textwidth]{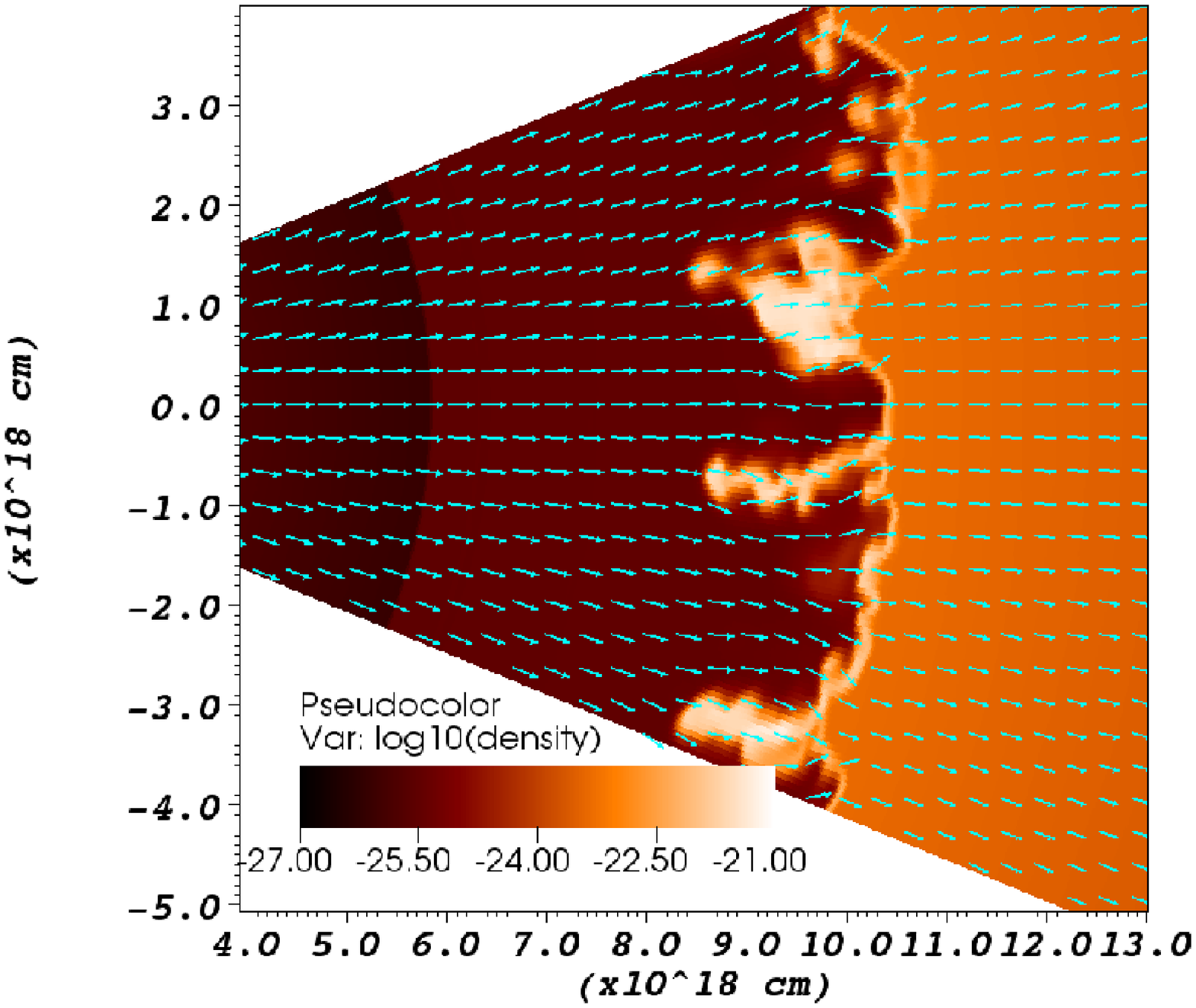}}}
}
\caption{3-D simulation the WR-RSG interaction in low resolution (A3). From left to right: density in cgs of the CSM after 7\,920 and 39\,200 years. 
This figure shows slices through the 3-D grid. These results closely resemble Fig.~\ref{fig:WR_RSG_2-D_low}, which has the same resolution in 2-D. 
However, the instabilities in the 3-D model are more pronounced than in 2-D and develop quicker.}
 \label{fig:WR_RSG_3-D_low}
\end{figure*}

\begin{figure*}
\FIG{
 \centering
\mbox{
\subfigure
{\includegraphics[width=0.5\textwidth]{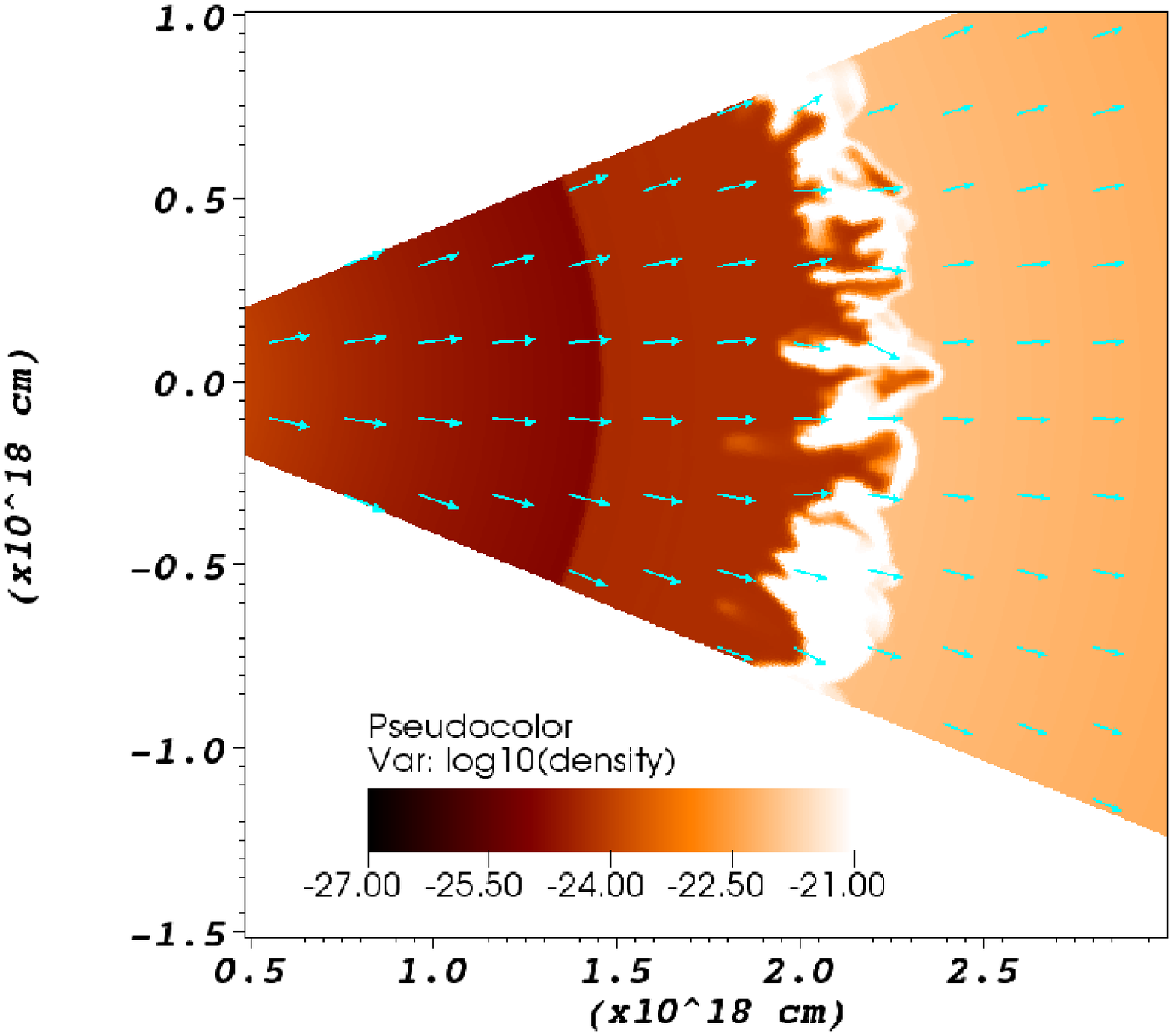}}
\subfigure
{\includegraphics[width=0.5\textwidth]{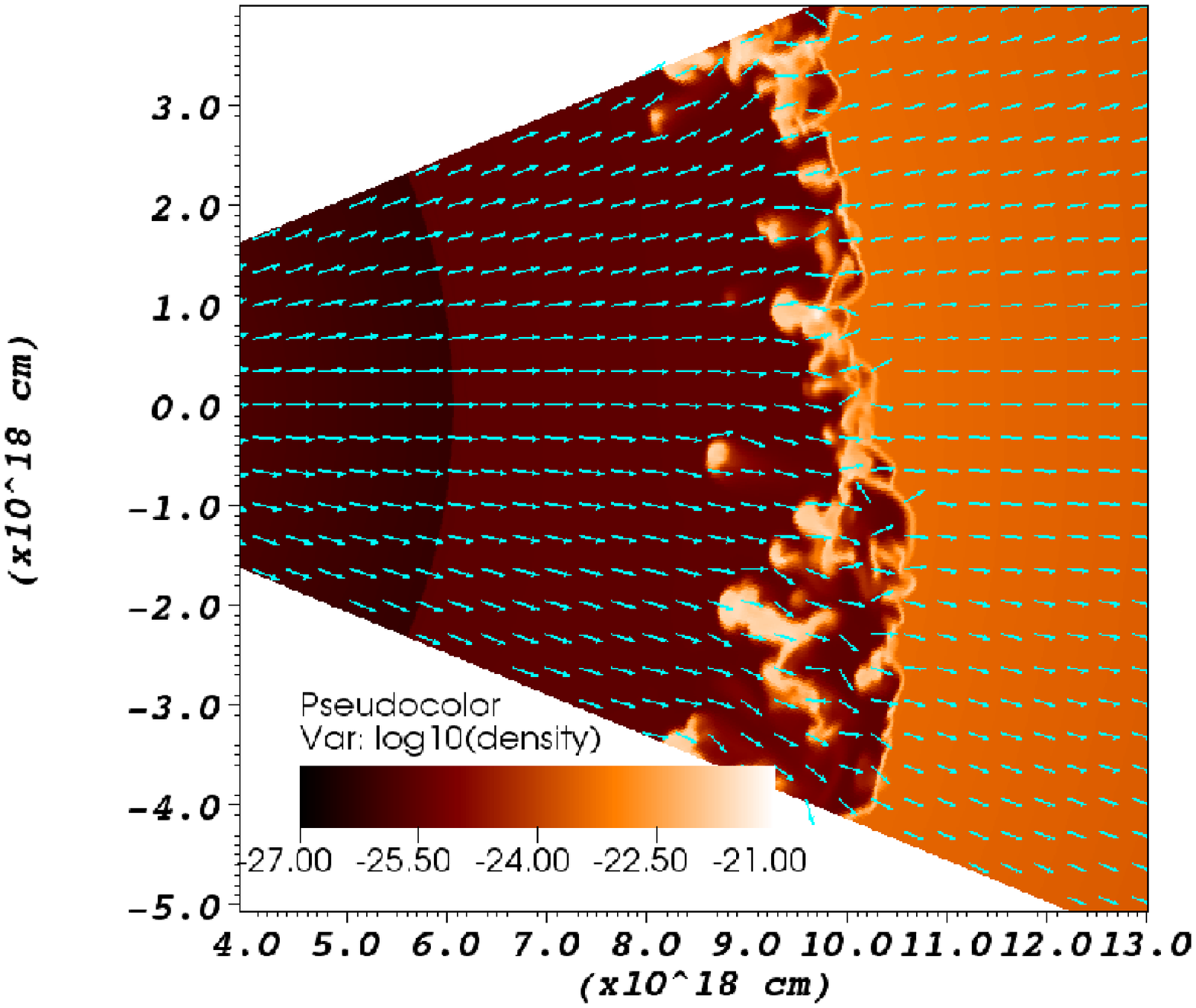}}}
}
\caption{Similar to Fig.~\ref{fig:WR_RSG_3-D_low}, but for simulation A4, which has twice the resolution. As for the low resolution model, the 
instabilities resemble those from the 2-D result, but appear to be further developed. 
In the later stages they show a strong tendency to break off from the shell.}
 \label{fig:WR_RSG_3-D_high}
\end{figure*}

\section{Results}
\label{sec-results}
\subsection{1-D models}
\label{sec-1-D}
The 1-D results, which are the starting points for the 2-D and 3-D simulations are shown in Figs.~\ref{fig:WR_RSG_1-D}-\ref{fig:WR_LBV_1-D}. 
They show the free-streaming WR wind, which starts from the central star, the shocked wind (starting at $1.6\times10^{17}$\,cm in Fig.~\ref{fig:WR_RSG_1-D} and $6.5\times10^{17}$\,cm in Fig.~\ref{fig:WR_LBV_1-D}), 
the swept up shell ($3\times10^{17}$\,cm in Fig.~\ref{fig:WR_RSG_1-D} and $9.5\times10^{17}$\,cm in Fig.~\ref{fig:WR_LBV_1-D}), 
and the remnant of the giant wind. 
The wind termination shock is near-adiabatic in both cases, causing a density jump of approximately a factor four. 
The forward shock (at the front of the shell) is clearly radiative, as the density increases by more than three orders of magnitude. 

The thermal pressure is very high in the shocked wind region, as it balances the ram pressure of the fast WR wind, 
and remains constant over the contact discontinuity (at about $2.9\times10^{17}$\,cm in Fig.~\ref{fig:WR_RSG_1-D} and $9.5\times10^{17}$\,cm in Fig.~\ref{fig:WR_LBV_1-D}), 
since the swept up shell is in pressure equilibrium with the shocked wind. 

In the unshocked winds ($R<1.6\times10^{17}$\,cm and $R>3\times10^{17}$\,cm in Fig.~\ref{fig:WR_RSG_1-D} and $R<6.4\times10^{17}$\,cm and $R>9.3\times10^{17}$\,cm in Fig.~\ref{fig:WR_LBV_1-D}) the thermal pressure is comparatively low 
and is set only by the density and the minimum temperature of 10\,000\,K. 
In the WR wind, the ram pressure (set by the wind velocity) is about four orders of magnitude higher than the thermal pressure, which means that the thermal pressure of the fast wind has no influence on the motion of the shell. 
The thermal pressure in the slow (RSG/LBV) wind could become significant, as it has to be compared with the forward motion of the shell relative to the slow wind, which is much slower than the fast wind velocity. 
Should the thermal pressure of the slow wind gas approach the ram pressure felt by the shell as it sweeps up the slow wind, then the thermal pressure could influence the motion of the shell.
As will be discussed in Section~\ref{sec-shellspeed}, the forward velocity of the shell is about 85\,$\kms$ for the WR-RSG interaction and about 300\,$\kms$ for the WR-LBV interaction. 
As a result, in the co-moving frame of reference of the shell, the ram pressure exerted by the slow wind on the shell (which is actually the effect of the shell moving into the slow wind) will be 50 and 800 times larger respectively, 
since the ram pressure increases with the velocity squared and the sound speed in the unshocked wind (which sets the thermal pressure) is about 12\,$\kms$. 
This means that the thermal pressure in the slow wind does not influence the motion of the shell significantly, which makes our assumption of complete ionization of hydrogen in the entire grid acceptable.

\subsection{The WR-RSG sequence}
In Fig.~\ref{fig:WR_RSG_2-D_low} we show the results from simulation~A1: A 2-D model of the expansion of a WR wind driven shell into the RSG wind with low resolution. 
At 7\,920 years (Fig.~\ref{fig:WR_RSG_2-D_low}, left panel), the shell already shows instabilities. 
These are thin-shell instabilities of the linear-Vishniac type \citep{Vishniac:1983}, caused by ram pressure from the unshocked RSG wind on the outside of the shell, 
balanced by the thermal pressure of the shocked WR wind. 
The instabilities are small compared to the radius of the shell. 
As the shell moves outward into the slow RSG wind, it forms large scale RT instabilities as well as Vishniac instabilities, 
as a result of the density difference between the shocked WR wind and the shell it drives. 
The size of these instabilities is on the same order of magnitude as the radius of the shell itself ($\sim\,0.5-1\,$pc). 
Figure.~\ref{fig:WR_RSG_2-D_low} right panel shows the density at 39\,200 years. 
Vishniac instabilities are still visible, although they are much smaller than the RT instabilities. 

The velocity field shows considerable deviation from the general radial motion, as the shocked WR wind flows around the RT instabilities. 
The large RT instability at the bottom (right panel of Fig.~\ref{fig:WR_RSG_2-D_low}) blocks the radial motion of the shocked wind material completely. 
This creates an ``empty'' region with lower density (and therefore lower thermal pressure) between the instability and the shell. 
The shocked wind material is pushed into this region by the thermal pressure gradient, resulting in a strong latitudinal motion. 
Because of the instabilities, there is motion in the latitudinal direction inside the shell itself as well. 

The high resolution run (simulation~A2) is shown in Fig.~\ref{fig:WR_RSG_2-D_high}. 
Although generally similar to simulation A1, the instabilities tend to be smaller (by about 50 percent) and show much more structure, which is understandable, 
since the smaller structures could not be resolved by the low resolution grid. 
The initial thin-shell instabilities have larger amplitudes than in the low resolution model, due to a shorter growth time (see Section~\ref{sec-instabilities}). 
Also, parts of the larger instabilities are breaking off and dissipating into the shocked WR wind, which does not occur in the low resolution simulation. 

The results of the low resolution 3-D simulation of the WR-RSG interaction (A3) are shown in Fig~\ref{fig:WR_RSG_3-D_low}. 
These have the same resolution as A1 (Fig.~\ref{fig:WR_RSG_2-D_low}, but on a 3-D grid. 
The figures show a 2-D slice through the 3-D grid, which can be compared directly to the 2-D results. 
Clearly, the 2-D and 3-D simulations produce qualitatively similar results, as the location of the shell and termination shock are nearly identical. 
In the first snapshot, the Vishniac instabilities in the 3-D model are larger than in the 2-D simulation and they appear to grow more quickly. 
It also shows the early stages of RT instabilities, which are much more developed then in the 2-D model.
In the later stage (Fig.~\ref{fig:WR_RSG_3-D_low}, right panel), the large scale RT instabilities dominate the morphology, 
which closely resembles the 2-D result. 
Their shape seems to be somewhat more longitudinal, but the difference is small.

The results of the high resolution run (A4) are shown in Fig.~\ref{fig:WR_RSG_3-D_high}. 
The instabilities seem to grow more quickly in the 3-D model than in the 2-D simulation (though the difference is less pronounced than for the low resolution model), 
with the early result (left panel of Fig.~\ref{fig:WR_RSG_3-D_high}) showing well developed Vishniac and RT instabilities. 
This is a direct result of the difference between 2-D and 3-D geometry. 
In a 2-D model, gas in the instability can only expand in the latitudinal direction. 
In the 3-D model it can move in the longitudinal direction as well, giving it an extra degree of freedom (See Sect.~\ref{sec-discussion}). 
The result after 39\,2000 years (right panel of Fig.~\ref{fig:WR_RSG_3-D_high}), resembles the 2-D result at similar resolution. 
However, the individual instabilities are smaller than in 2-D and have a strong tendency to detach from the shell.

\begin{figure*}
\FIG{
 \centering
\mbox{
\subfigure
{\includegraphics[width=0.5\textwidth]{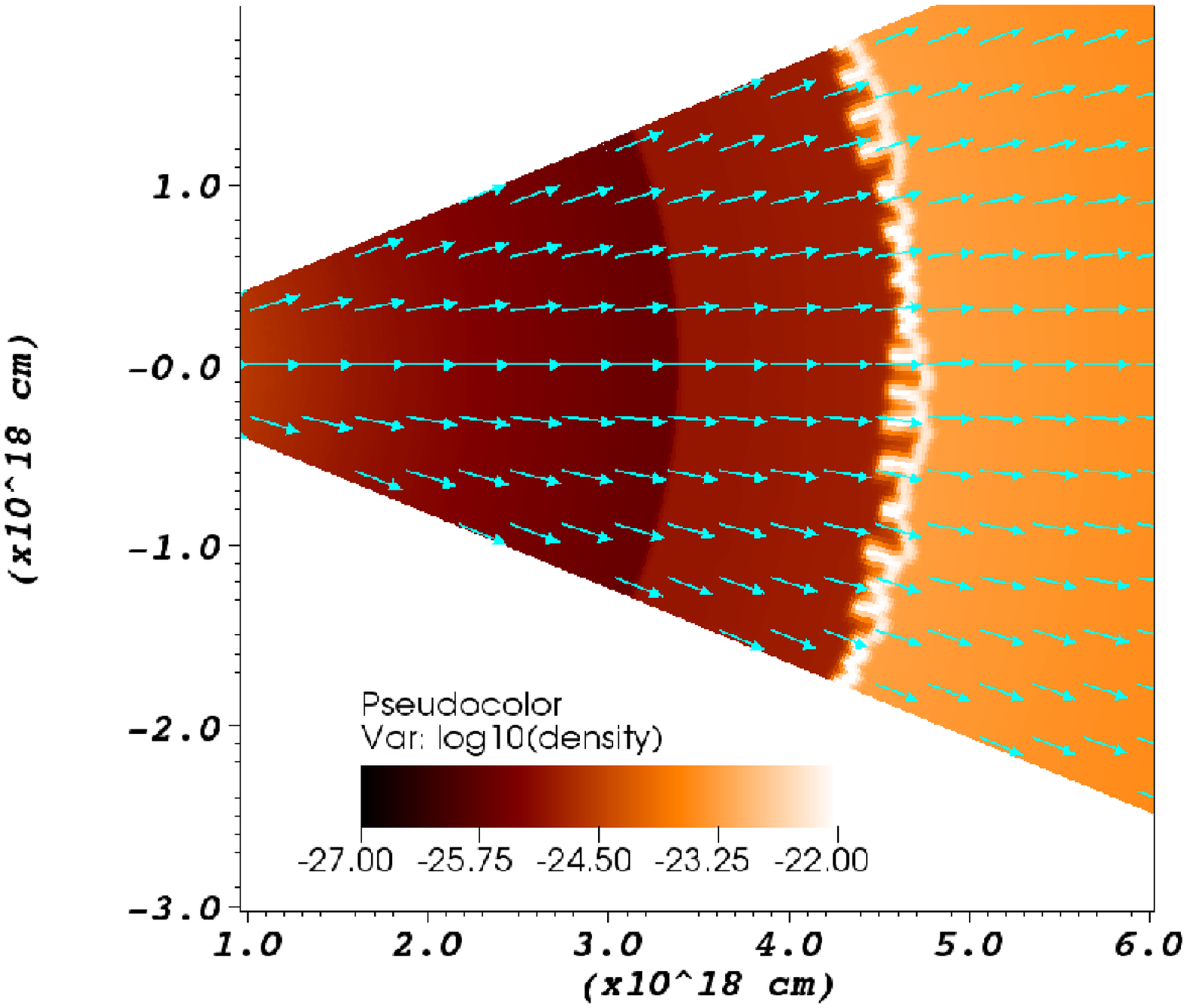}}
\subfigure
{\includegraphics[width=0.5\textwidth]{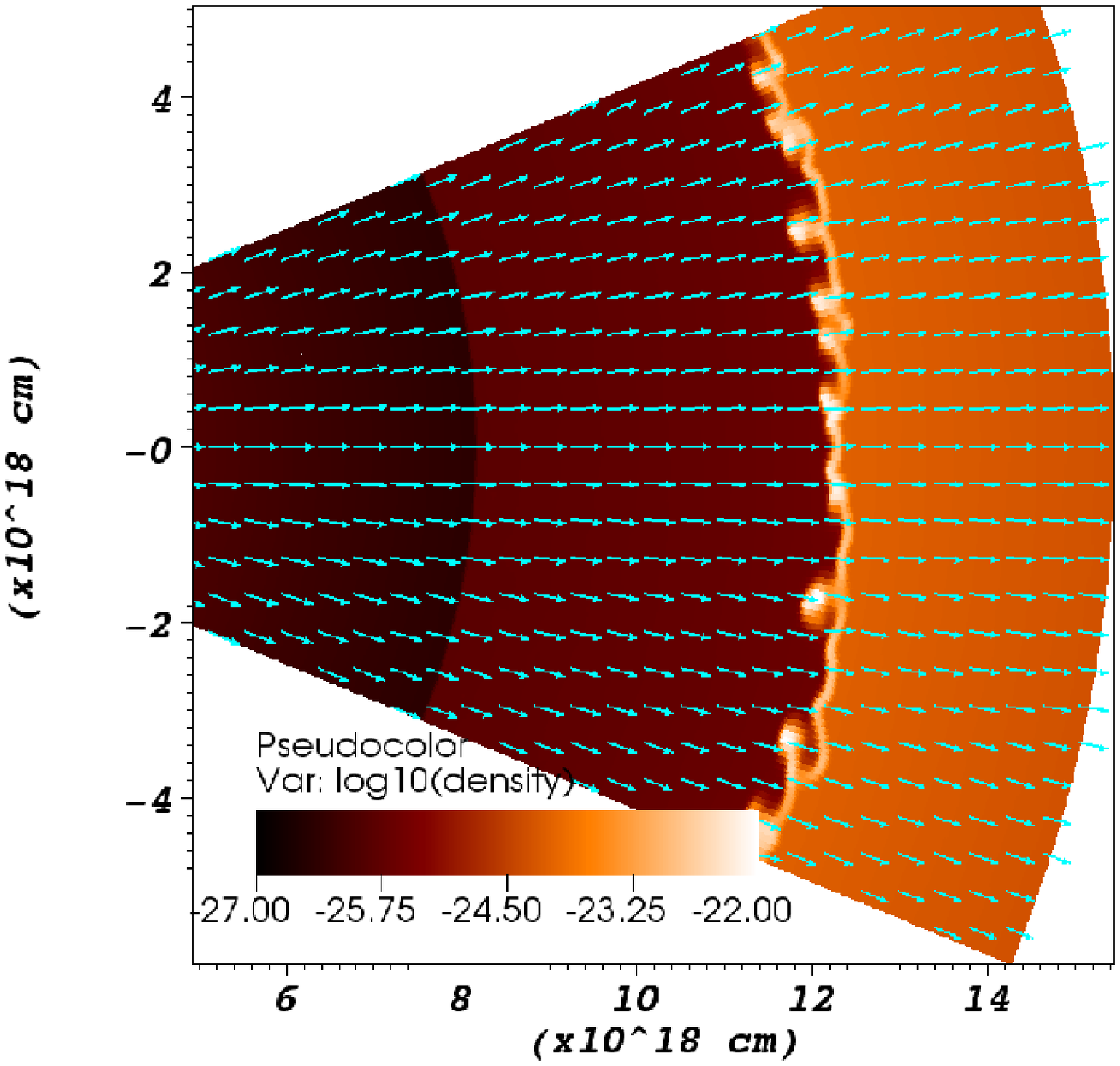}}}
}
\caption{The WR-LBV interaction in low resolution. 
This figure shows the density in cgs. as well as the velocity field for simulation B1 after 4\,010 and 11\,800~years. 
In the left panel, the shell mainly shows small Vishniac instabilities. 
In the right panel, some evidence of RT instabilities is visible, 
but unlike for the WR-RSG interaction (Figs.~\ref{fig:WR_RSG_2-D_low}-\ref{fig:WR_RSG_2-D_high}), they are small compared to the size of the shell.}
 \label{fig:WR_LBV_2-D_low}
\end{figure*}

\begin{figure*}
\FIG{
 \centering
\mbox{
\subfigure
{\includegraphics[width=0.5\textwidth]{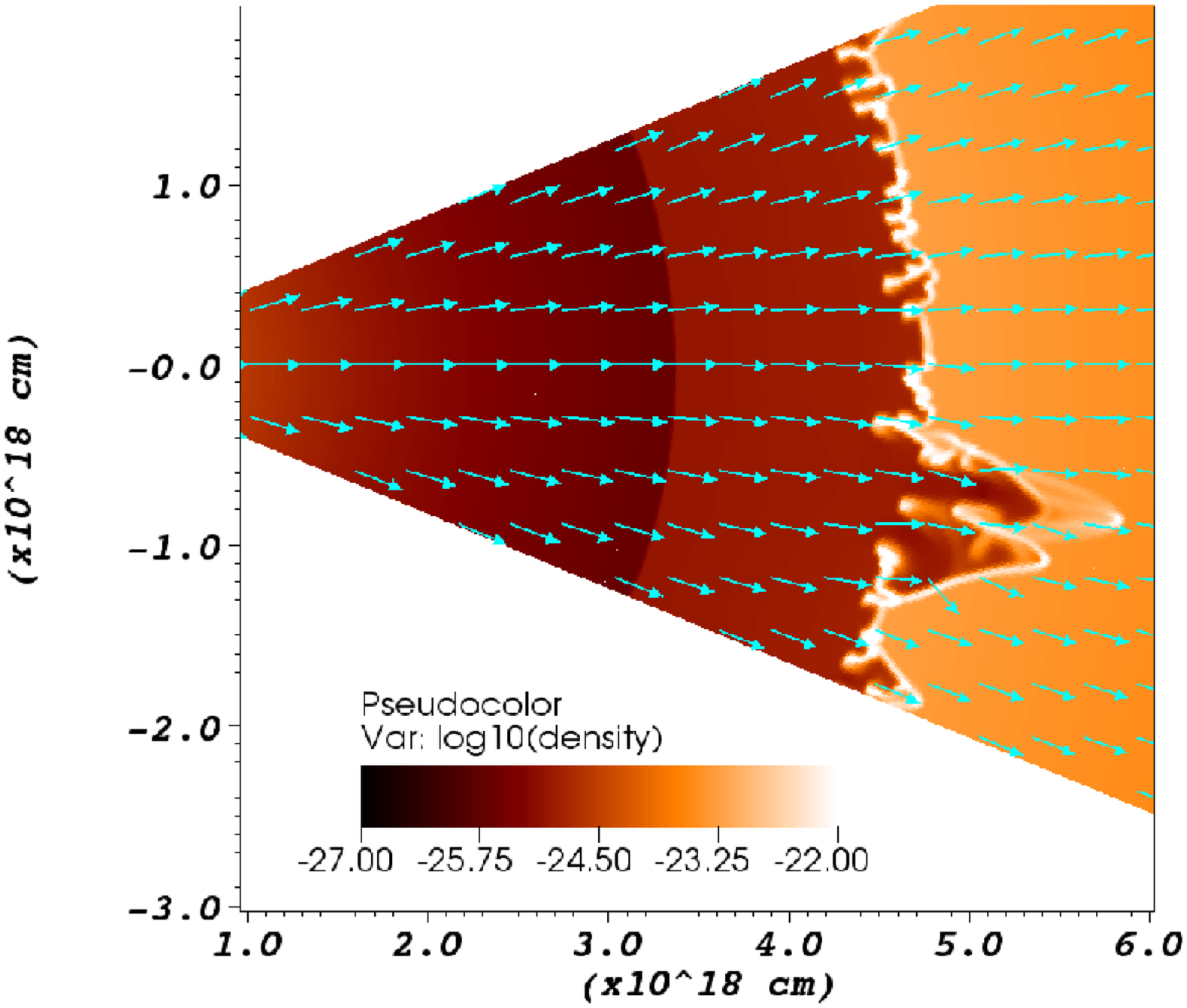}}
\subfigure
{\includegraphics[width=0.5\textwidth]{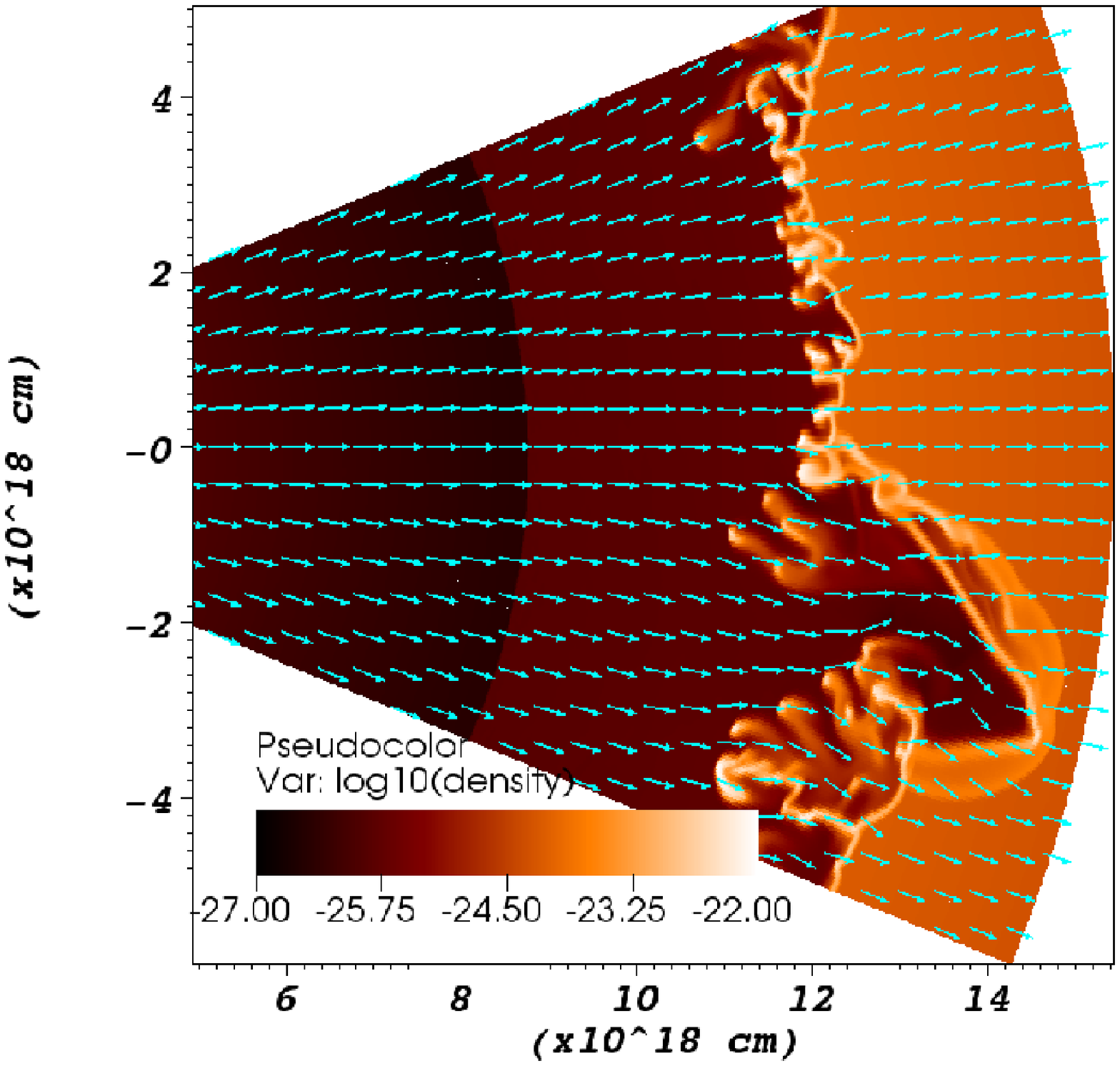}}}
}
\caption{Similar to Fig.~\ref{fig:WR_LBV_2-D_low}, but for high resolution (simulation~B2). 
Not only do the RT instabilities develop much further than in the low-resolution model, but one of the Vishniac instabilities grows to such a scale, 
that it deforms the general shape of the shell, with the hot gas of the shocked WR wind breaking out of the shell. 
This does not occur in the low resolution model.} 
 \label{fig:WR_LBV_2-D_high}
\end{figure*}

\begin{figure*}
\FIG{
 \centering
\mbox{
\subfigure
{\includegraphics[width=0.5\textwidth]{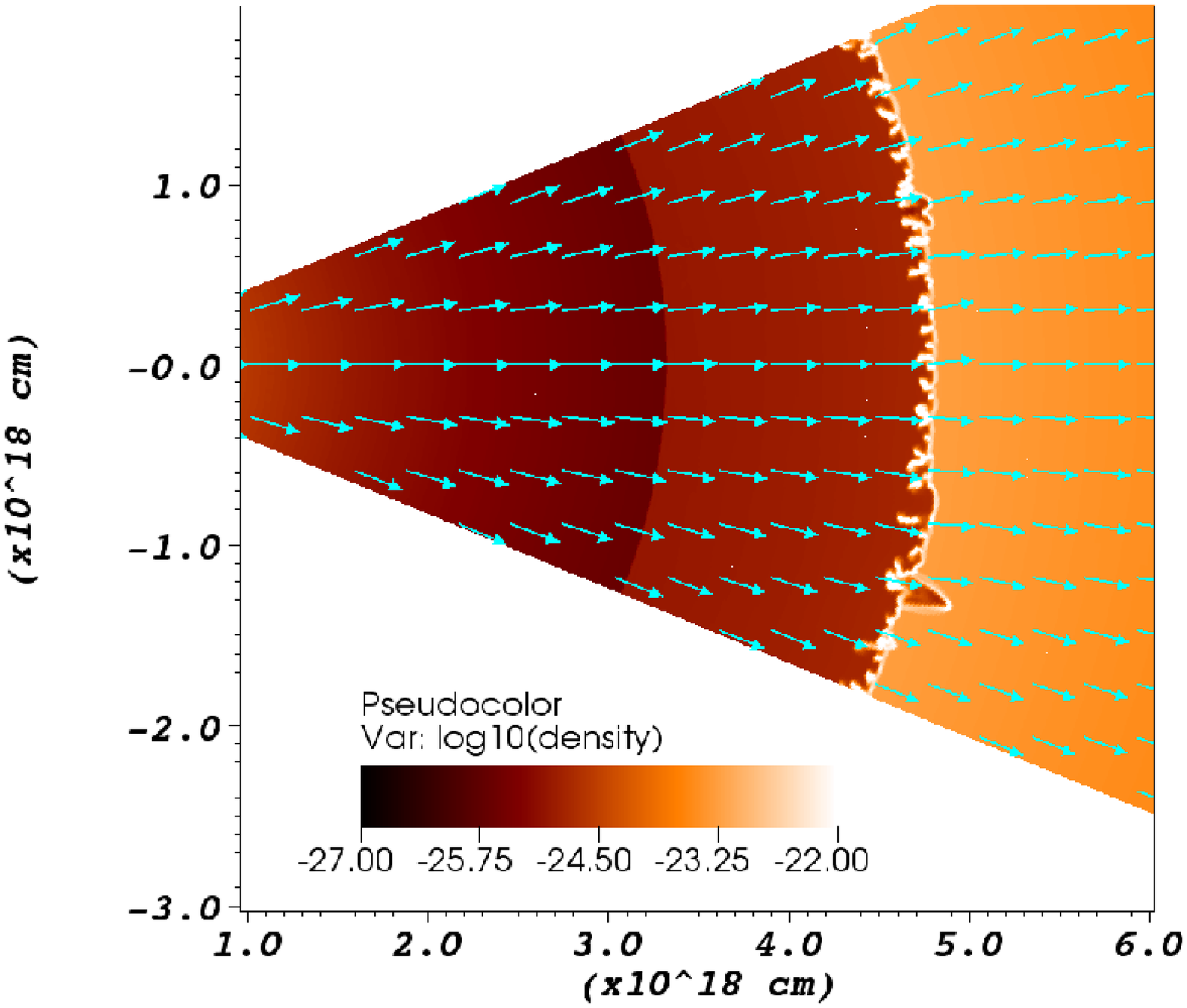}}
\subfigure
{\includegraphics[width=0.5\textwidth]{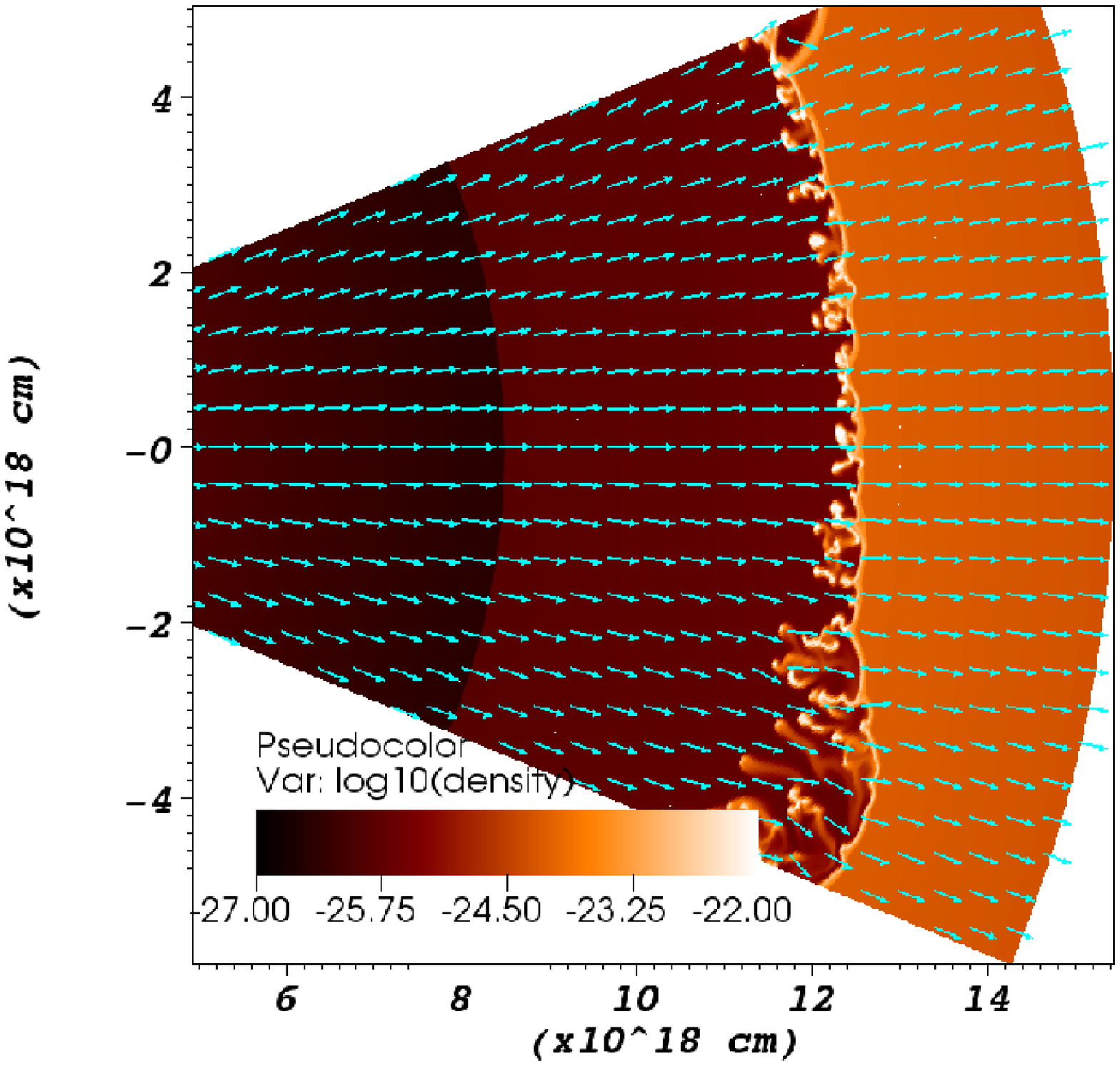}}}
}
\caption{Similar to Fig.~\ref{fig:WR_LBV_2-D_high}, but with double the resolution. 
A break-out similar to the one shown in Fig.~\ref{fig:WR_LBV_2-D_high} occurs in this simulation, 
but it does not grow to the same size and disappears over time.}
 \label{fig:WR_LBV_2-D_extrahigh}
\end{figure*}

\begin{figure*}
\FIG{
 \centering
\mbox{
\subfigure
{\includegraphics[width=0.5\textwidth]{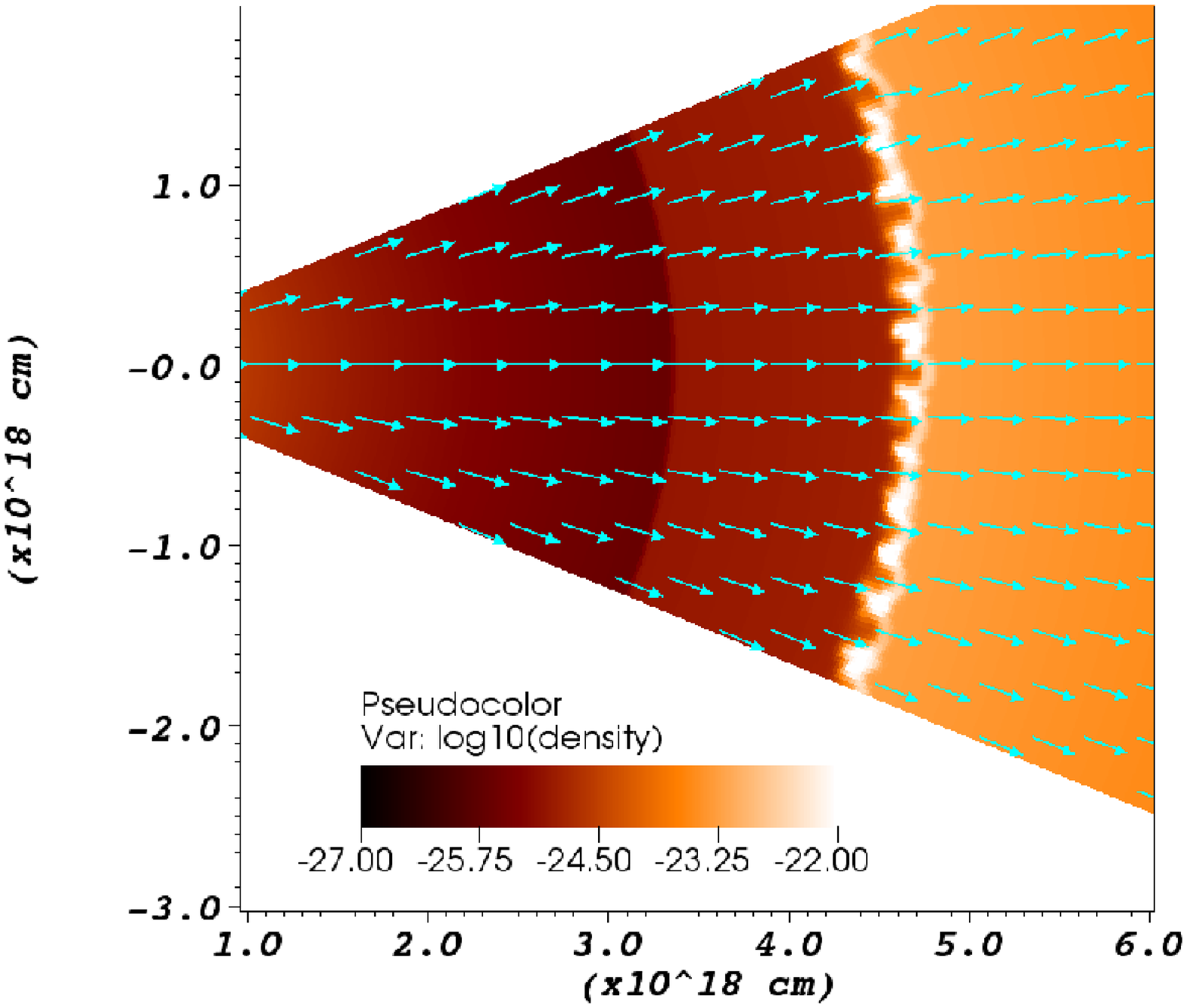}}
\subfigure
{\includegraphics[width=0.5\textwidth]{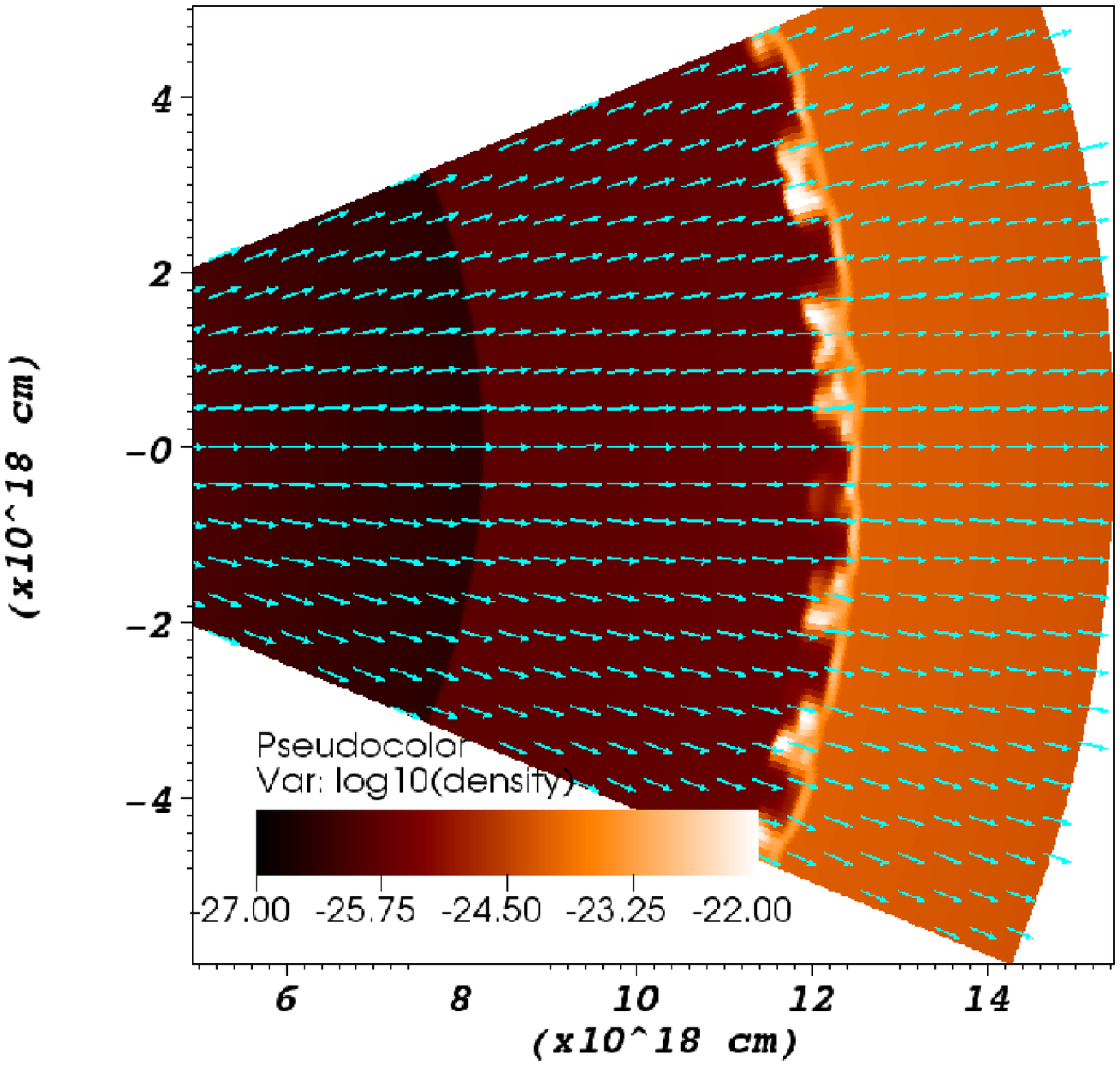}}}
}
\caption{Slices through the 3-D simulation of the WR-LBV interaction in low resolution (B3). From left to right: density in cgs of the CSM at 4\,010, 7\,920 and 11\,800~years. 
These results closely resemble Fig.~\ref{fig:WR_LBV_2-D_low}, which has the same resolution in 2-D.}
 \label{fig:WR_LBV_3-D_low}
\end{figure*}

\begin{figure*}
\FIG{
 \centering
\mbox{
\subfigure
{\includegraphics[width=0.5\textwidth]{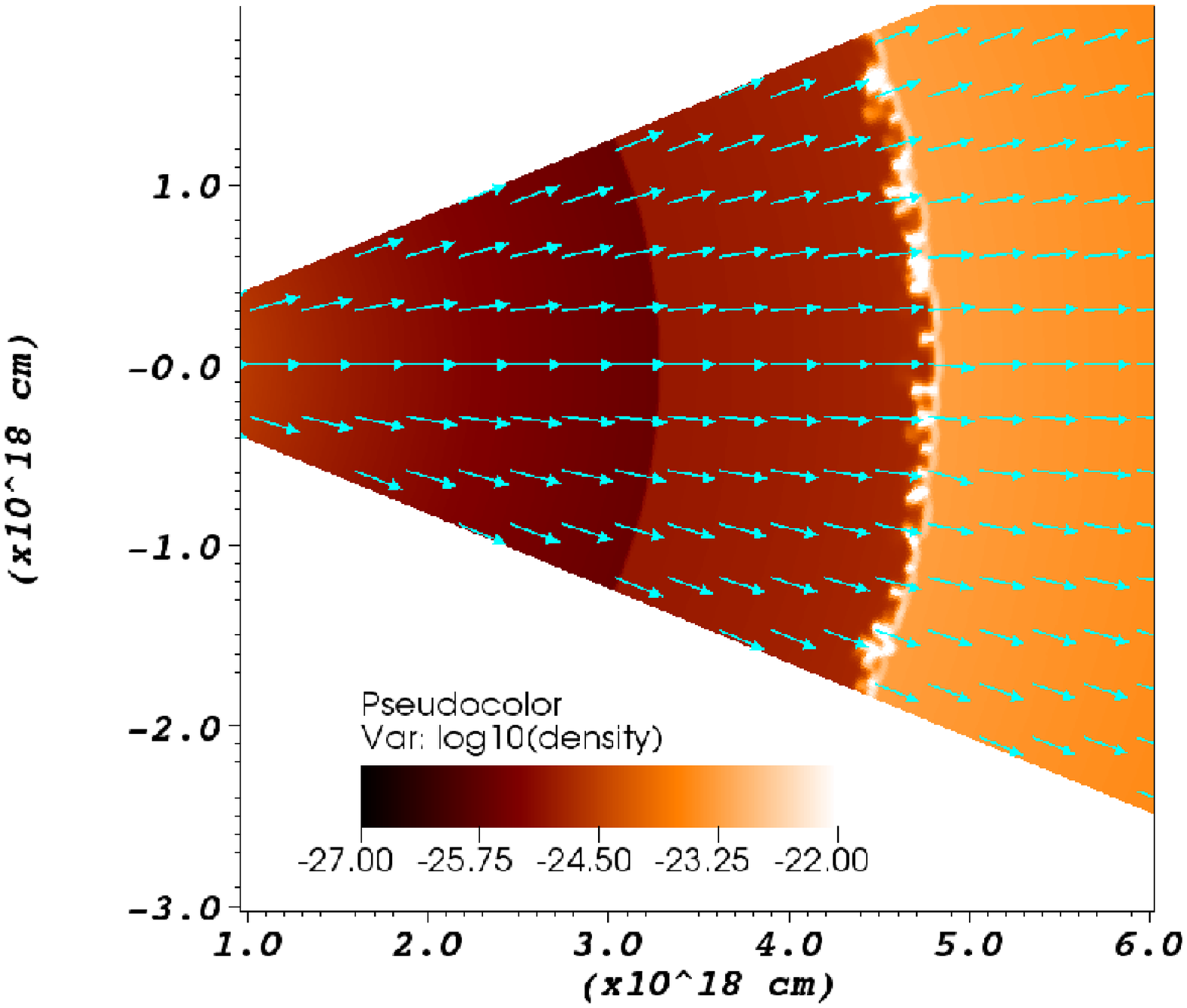}}
\subfigure
{\includegraphics[width=0.5\textwidth]{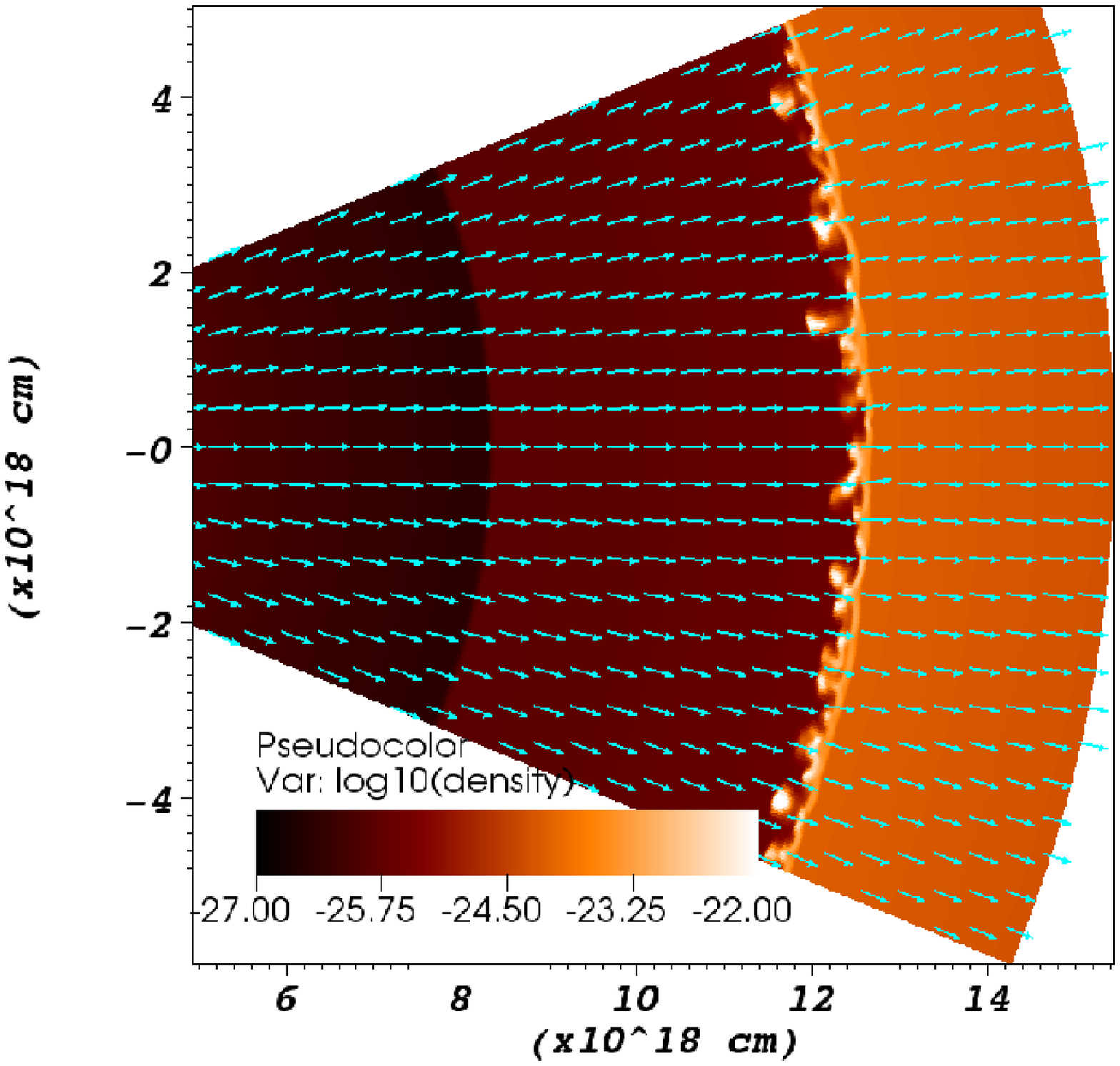}}}   
}
      \caption{Similar to Fig.~\ref{fig:WR_LBV_3-D_low}, but for simulation B4, which shows the same interaction in high resolution. 
Compared to the 2-D model at the same resolution (simulation B2, Fig.~\ref{fig:WR_LBV_2-D_high}), 
these results stand out due to the absence of any large instability.}
 \label{fig:WR_LBV_3-D_high}
\end{figure*}

\subsection{The WR-LBV sequence} 
Figure~\ref{fig:WR_LBV_2-D_low} shows the low resolution result for the WR-LBV interaction in simulation B1 after 4\,010 and 11\,800~years. 
Unlike the WR-RSG interaction, the instabilities remain small ($\lesssim0.1\,$pc), although both linear-Vishniac and RT instabilities occur. 
There are several causes for this effect. 
First, despite its higher massloss rate, the density in the LBV wind is lower than in the RSG wind, due to its higher velocity ($\rho\propto v^{-1}$). 
As a result, the density difference between the WR wind and the LBV wind is smaller than between the WR wind and the RSG wind. 
This also influences the size of the RT instabilities, which depends on the density contrast. 
Second, the shell travels much faster ($\sim330\,\kms$ vs. $\sim85\,\kms$), giving the instabilities much less time to grow. 
The flow is typically radial throughout the grid, with only small deviations around the instabilities.

The high resolution WR-LBV simulation (simulation~B2, Fig.~\ref{fig:WR_LBV_2-D_high}) shows considerable difference with the low resolution run. 
The low resolution run clearly has trouble resolving the instabilities, which appear to be identical as they follow the shape of the gridcells. 
The high resolution model resolves the instabilities properly so they show local variation.
In the high resolution model, one of the Vishniac instabilities grows to such a size that it breaks up the spherical symmetry of the shell 
and causes a turbulent flow pattern. 
Apart form this large instability, the flow is almost completely radial, with only small local variations. 
The RT instabilities remain thin and don't grow into the large clumps that occur in the WR-RSG simulations. 
This is due to the much faster forward motion of the shell, which does not give the instabilities time to expand 
and the lower density contrast between the shell and the shocked wind region.  

The evolution of the large instability deviates considerably from that of the shell in general. 
Its radial expansion is much faster and the small RT instabilities at the base tend to block part of the flow from the shocked wind region. 
As a result, the pressure driving the instability outwards becomes anisotropic and it starts to take on some of the characteristics of a non-linear Vishniac instability \citep{Vishniac:1994}. 
The instability never fully transitions to the non-linear regime because the thermal pressure in the shocked wind region remains larger than the ram-pressure. 
The nature of the forward shock of the instability also changes. 
Due to its high radial velocity shock-heating increases and the shock becomes partially adiabatic. 
This leads to an increase in the thickness of the shell (Fig.~\ref{fig:WR_LBV_2-D_high}, right panel), 
which stops the evolution of the Vishniac instability as the curvature of the forward shock starts to decrease relative to the triangular shape of the contact discontinuity. 

In order to investigate whether the large distortion of the shell that occurs in simulation~B2 is representative for higher resolution solutions, 
we repeat the simulation with one more level of grid refinement, which gives us an effective resolution of $3200\times512$ gridpoints. 
It is to be noted that the random density variations in fact lead to different realizations (on top of the resolution differences), preventing us to reach true numerically converged states.
The higher resolution result is shown in Fig.~\ref{fig:WR_LBV_2-D_extrahigh}. 
Although a large scale instability does form in the earlier stages of the evolution of the shell, it does not grow to the same scale as in Fig.~\ref{fig:WR_LBV_2-D_high}. 
Moreover, the large instability spreads out in the latitudinal direction, rather than retain its triangular shape. 
The small scale instabilities are similar to those that occur in simulation~B2.
We therefore speculate that the large-scale deformations from simulation~B2 represent rare events.

Figures~\ref{fig:WR_LBV_3-D_low} and \ref{fig:WR_LBV_3-D_high} show slices through the 3-D simulations of the WR-LBV interaction (simulations~B3 and B4). 
Both simulations mainly resemble the low resolution 2-D model (simulation B1, Fig.~\ref{fig:WR_LBV_2-D_low}). 
The low resolution 3-D model shows more variation in the shape of the instabilities than the 2-D model.
Comparisons between the high resolution models are difficult due to the absence of the single large instability that occurred in the high resolution 2-D model (Fig.~\ref{fig:WR_LBV_2-D_high}). 
Since it does not occur anywhere in the 3-D models (See also Figs.~\ref{fig:WR_LBV_low_angle} and \ref{fig:WR_LBV_high_angle}), 
such large scale structures are probably rare, though a ``blow-out'' has been observed in, for example, \object{RCW~58}. 
A comparison between simulation between simulation~B4 and the extra-high resolution model shown in Fig.~\ref{fig:WR_LBV_2-D_extrahigh} shows that the results are very much alike, though, 
once again, the extra resolution of the latter helps to resolve the instabilities. 
As in 2-D, the 3-D simulation shows more structured instabilities at high resolution, which have a tendency to break off from the shell. 
The instabilities seem to have a tendency to cluster together in small groups. 
These groups are the result of the filamentary 3-D structure of the instabilities (See Sect.~\ref{sec-observation}) and denote the places where the slice intersects 
with the knots between filaments.

\begin{figure*}
\FIG{
 \centering
\mbox{
\subfigure
{\includegraphics[width=0.40\textwidth,angle=-90]{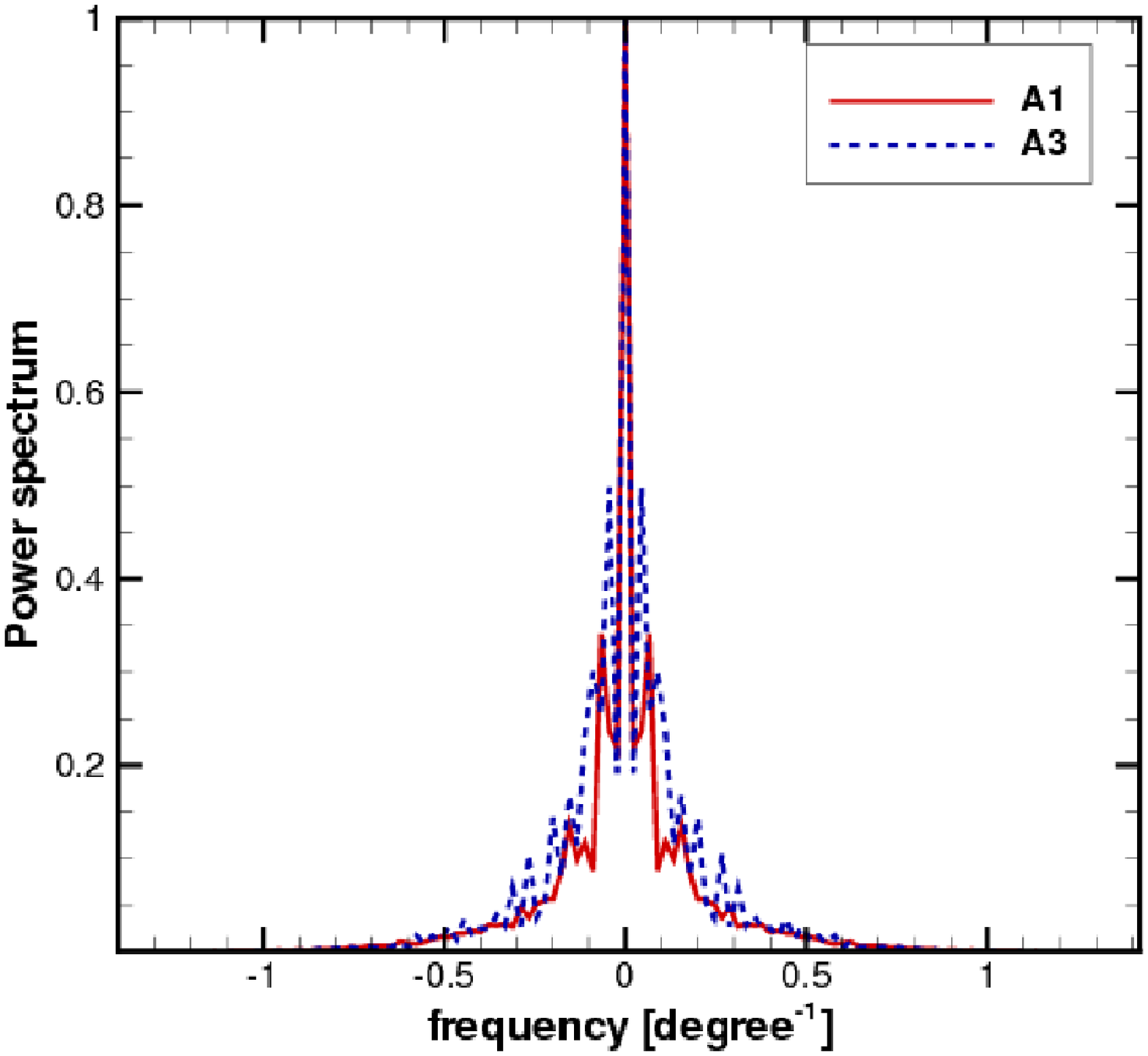}}
\subfigure
{\includegraphics[width=0.40\textwidth,angle=-90]{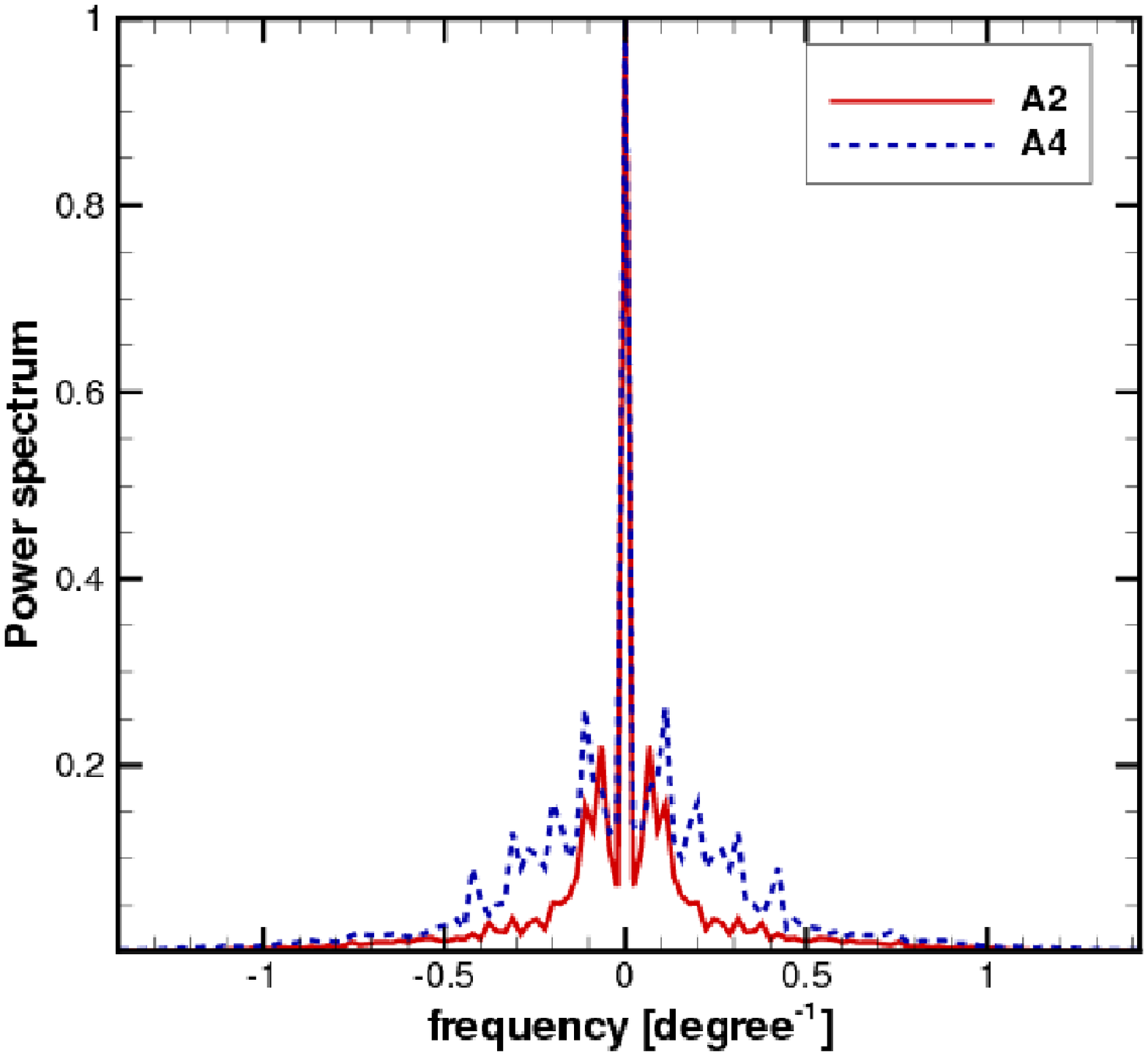}}}
}
\caption{This figure shows the power spectrum calculated from a 1-D Fourier transform along the latitudinal axis 
for the final results of simulations A1 and A3, being 2-D vs. 3-D (left panel) and their equivalents at higher resolution, A2 and A4(right panel). 
All simulations show a peak at about 0.8 per latitudinal degree, corresponding to the large RT instabilities. 
The higher resolution simulations (A2 and A4) show more high order maxima than the low resolution models. 
Also, the peaks corresponding to instabilities tend to be higher with respect to the zero-th order peak for the 3-D simulations than for the 2-D models}
 \label{fig:WR_RSG_fft}
\end{figure*}

\begin{figure*}
\FIG{
 \centering
\mbox{
\subfigure
{\includegraphics[width=0.40\textwidth,angle=-90]{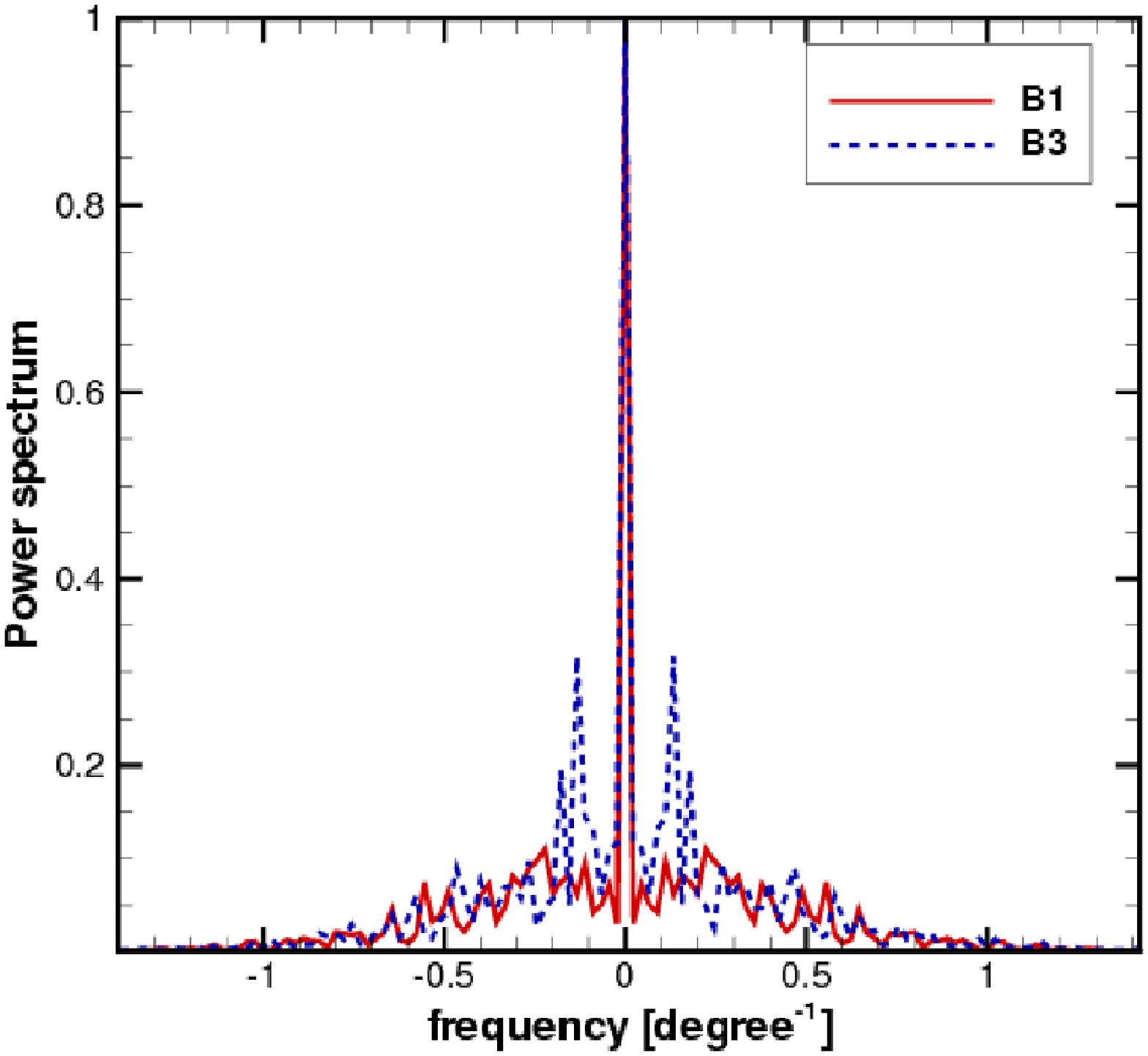}}
\subfigure
{\includegraphics[width=0.40\textwidth,angle=-90]{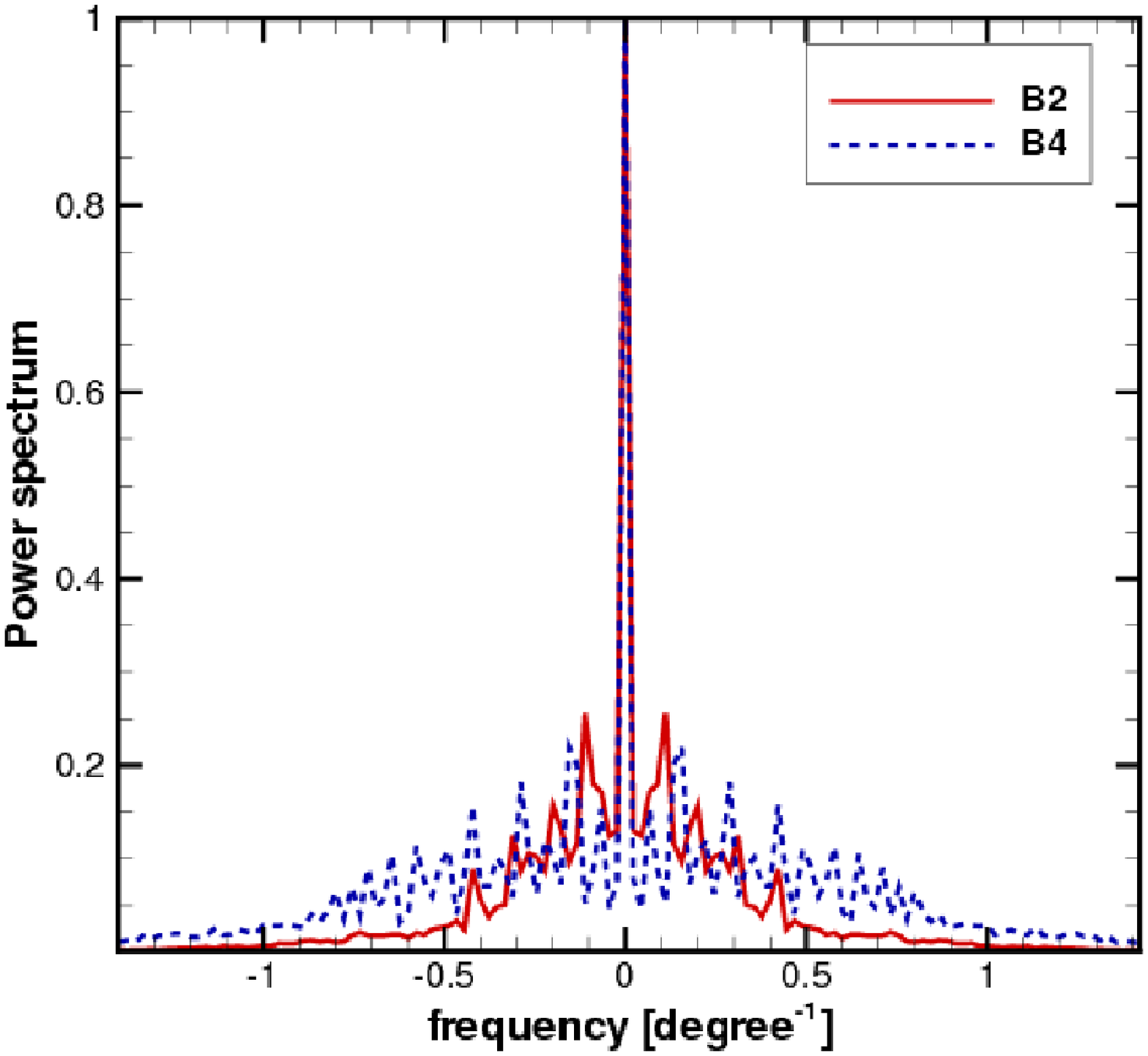}}}
}
\caption{Similar to Fig.~\ref{fig:WR_RSG_fft}, but for the final results of simulations~B1 and B3 (left panel) and B2 and B4 (right panel). 
As for the WR-RSG interaction, the higher resolution models tend to show more higher order maxima and the 
3-D models have higher peaks corresponding to the instabilities. 
The power spectrum of simulation B3 shows a peak at about 0.05 per latitudinal angle, which corresponds to the single large instability.}
 \label{fig:WR_LBV_fft}
\end{figure*}

\section{Instabilities}
\label{sec-instabilities}
There are several different mechanisms for instabilities at work in our hydrodynamical models of circumstellar shells.
Initially, in all models, we see linear thin-shell, or linear Vishniac, instabilities \citep{Vishniac:1983}. 
These occur whenever a thin shell is compressed between thermal pressure on one side and ram pressure on the other as is the case in all our simulations. 
They are most visible in the earlier stages of the shell formation because of their short growth time, which is comparable to the sound crossing time of the shell \citep[Eqn.~2.23 from][]{Vishniac:1983}. 
Since the initial shell has a cross-section of less than 1/10$^{th}$ of a parsec and the sound speed in the shell (at 10\,000~K) is about 12\,$\kms$, 
this gives us a formation time of a few thousand years, 
short enough to be visible in the first image (left panels in Figs.~\ref{fig:WR_RSG_2-D_low}-\ref{fig:WR_LBV_2-D_high} and  \ref{fig:WR_RSG_3-D_low}-\ref{fig:WR_LBV_3-D_high}.) 

The non-linear variant of the thin-shell instability \citep{Vishniac:1994} could theoretically occur if the shell was subjected to ram pressure from both sides. 
For this to happen, the wind termination shock would have to be radiative, which is possible at high densities. 
Because the density of the wind is high close to the star, all wind termination shocks are initially radiative. 
However, this is only for a very short period of time, as the wind termination shock has already become adiabatic after 100\,years in the 1-D models (see Section~\ref{sec-1-D}). 
This does not give the non-linear thin-shell instabilities enough time to grow. 

In the later stages of their evolution, the shells are dominated by RT instabilities. 
These result from a situation where a low-density gas (the shocked WR wind) accelerates a much denser gas (the RSG or LBV wind).
The size of these instabilities is much larger than for the thin-shell instabilities, 
and in the case of the WR-RSG interaction they grow to such a scale that they can deform the overall shape of the shell. 

The third form of instability is the radiative cooling instability. 
This is caused by the strong density dependence of the radiative cooling ($\propto\rho^2$). 
As a result, high density regions cool much quicker than low density regions leaving them with a lower thermal pressure. 
This causes them to be compressed by the surrounding medium, which in turn increases their density. 
Typically, such instabilities occur on a small scale, which makes them difficult to resolve, leading to numerical problems; especially in 2-D simulations \citep{vanMarleKeppens:2011}.
They cause local density differences in the fast cooling shell, which can serve as a starting point for the larger thin-shell and RT instabilities. 
We limit the growth of radiative cooling instabilities by maintaining a minimum temperature of 10\,000~K. 
This stops the cooling clumps from being compressed further, once they reach this temperature. 

A fourth instability, which cannot occur in our models because we neglect detailed radiative transfer, is the photo-ionization instability. 
This effect is observed when only part of the gas is photo-ionized. 
The ionized gas, which is much hotter than the neutral gas will tend to expand, leaving it with a lower density, which in turn is easier to photo-ionize. 
The neutral gas, which is compressed, becomes denser, recombining more quickly and becoming impossible for the ionizing photons to penetrate. 
This effect can be important in the structure of planetary nebulae \citep{Garcia-Seguraetal:1999}. 
This can also become important for WR nebulae. 
Models by \citet{ToalaArthur:2011}, which include photo-ionization through radiative transfer show that, though the temperature of the gas in the shells is typically at about 10\,000~K, 
indicating that our approximation of complete ionization is reasonable, the denser clumps are only partially ionized and cool to a lower temperature.

\subsection{Powerspectra of the instabilities in circumstellar shells}
To make a quantitative comparison between the 2D and 3D results, we calculate the Fourier transform of the particle density along the latitudinal grid axis 
for the final results 
of each simulation 2-D simulation (right panels of Figs.~\ref{fig:WR_RSG_2-D_low}-\ref{fig:WR_RSG_2-D_high} and \ref{fig:WR_LBV_2-D_low}-\ref{fig:WR_LBV_2-D_high}) as well as the slices through the final results of the 3-D simulations 
(right panels of Figs.~\ref{fig:WR_RSG_3-D_low}-\ref{fig:WR_RSG_3-D_high} and \ref{fig:WR_LBV_3-D_low}-\ref{fig:WR_LBV_3-D_high}). 
This allows us to identify the recurring pattern of the instabilities as well as their wavelength. 

The resulting power spectra for the WR-RSG interaction, normalized to the peak height 
of the zero-th order maximum are shown in Fig.~\ref{fig:WR_RSG_fft} 
(for details of the Fourier quantification, see Appendix~\ref{sec-fft}). 
Apart from the zero-th order frequency peak, the result of simulation A1 (left panel of Fig.~\ref{fig:WR_RSG_fft}) shows a peak at about 0.08 per degree latitude. 
This corresponds to the large scale RT instabilities of which there are four over the 45 degree angle of the simulation. 
A second peak, though very faint appears at about 0.15 per latitudinal degree.
The 3-D simulation (A3) shows a similar pattern. 
The peaks lie slightly closer to the zero-th order maximum, which is to be expected since the simulation only shows three major RT instabilities. 
The peaks corresponding to these instabilities are much higher compared to 2-D simulation. 
This can be explained by the fact that the RT instabilities are longer and thinner, so the pattern repeats itself over a larger part of the radial domain. 
The Fourier spectra for simulations A2 and A4 (right panel of Fig.~\ref{fig:WR_RSG_fft}) shows a pattern similar to the low resolution models. 
In this case, the frequency of the instabilities is higher for the 3-D model than for 2-D. 
The 3-D model now shows several higher order peaks, which are much higher compared to the zero-th maximum. 
As for the low resolution models, the repetitive pattern of the instabilities is more pronounced in 3-D than in 2-D. 

The results for the WR-LBV interaction are shown in Fig.~\ref{fig:WR_LBV_fft}. 
Both 2-D simulations show peaks at a frequency of about 0.3 per latitudinal angle, corresponding to about 12 waves over the $45^o$ angle of the domain. 
This frequency corresponds with the instabilities noted in Sect.~\ref{sec-results}. 
A second peak at about 0.5 per latitudinal angle is faintly visible for simulation B1 and very clearly for simulation B2. 
This corresponds to the smaller subdivisions of the initial instabilities, which are clearly visible for simulation B2. 
Simulation B2 also shows a peak at 0.05 per latitudinal angle. 
This is equal to about 2  waves over the $45^o$ degree domain, which corresponds to the shape of the large Vishniac instability. 
The 3-D simulations show a different pattern. 
Both have maxima at about 0.1-0.15 per latitudinal angle and a series of peaks at higher frequencies. 
The peaks at about 0.1-0.15 are especially pronounced for the low resolution simulation (B3). 
This corresponds to approximately 5 waves over the $45^o$ degree domain. 
A close inspection of the density plot (right panel of Fig.~\ref{fig:WR_LBV_3-D_low}) shows that the small RT fingers tend to cluster together 
where the slices intersect with the knots in the filamentary structure of the shell (See Sect.~\ref{sec-observation}), forming 4-5 groups. 
This pattern is less clear for the higher resolution model (B4), although the tendency of the fingers to appear close together is still visible. 

A similar analysis could be done on observational results to identify the patterns of the instabilities and compare them to observational results.

   \begin{figure*}
\FIG{
   \centering
   \includegraphics[width=0.95\textwidth]{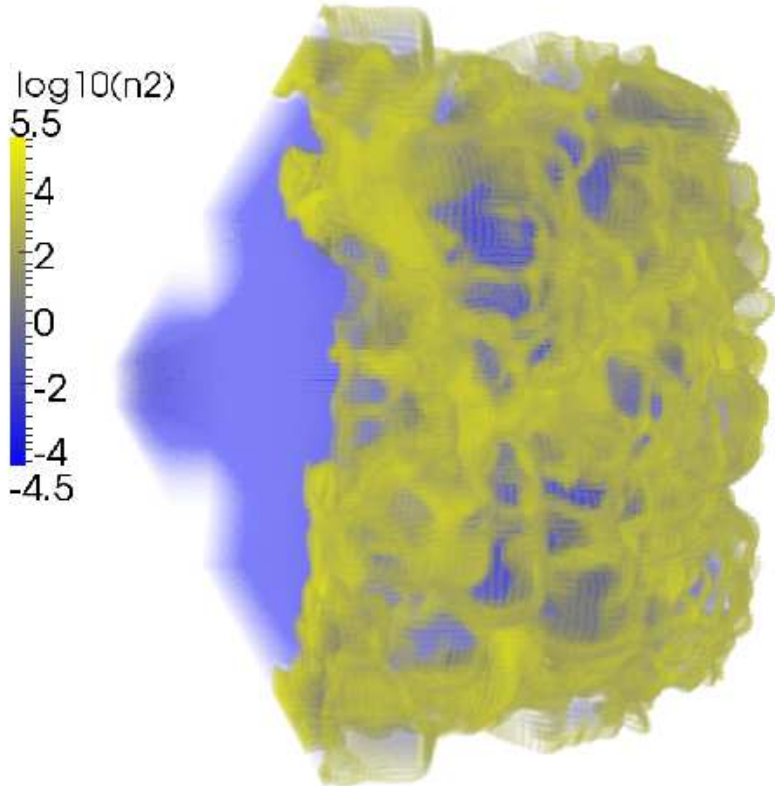}
}
      \caption{Particle density squared [1/cm$^6$] volume rendering of the low resolution WR-RSG interaction (simulation~A3) at the same time as the last panel of  Fig.~\ref{fig:WR_RSG_3-D_low}. }
         \label{fig:WR_RSG_low_angle}
   \end{figure*}

   \begin{figure*}
\FIG{
   \centering
   \includegraphics[width=0.95\textwidth]{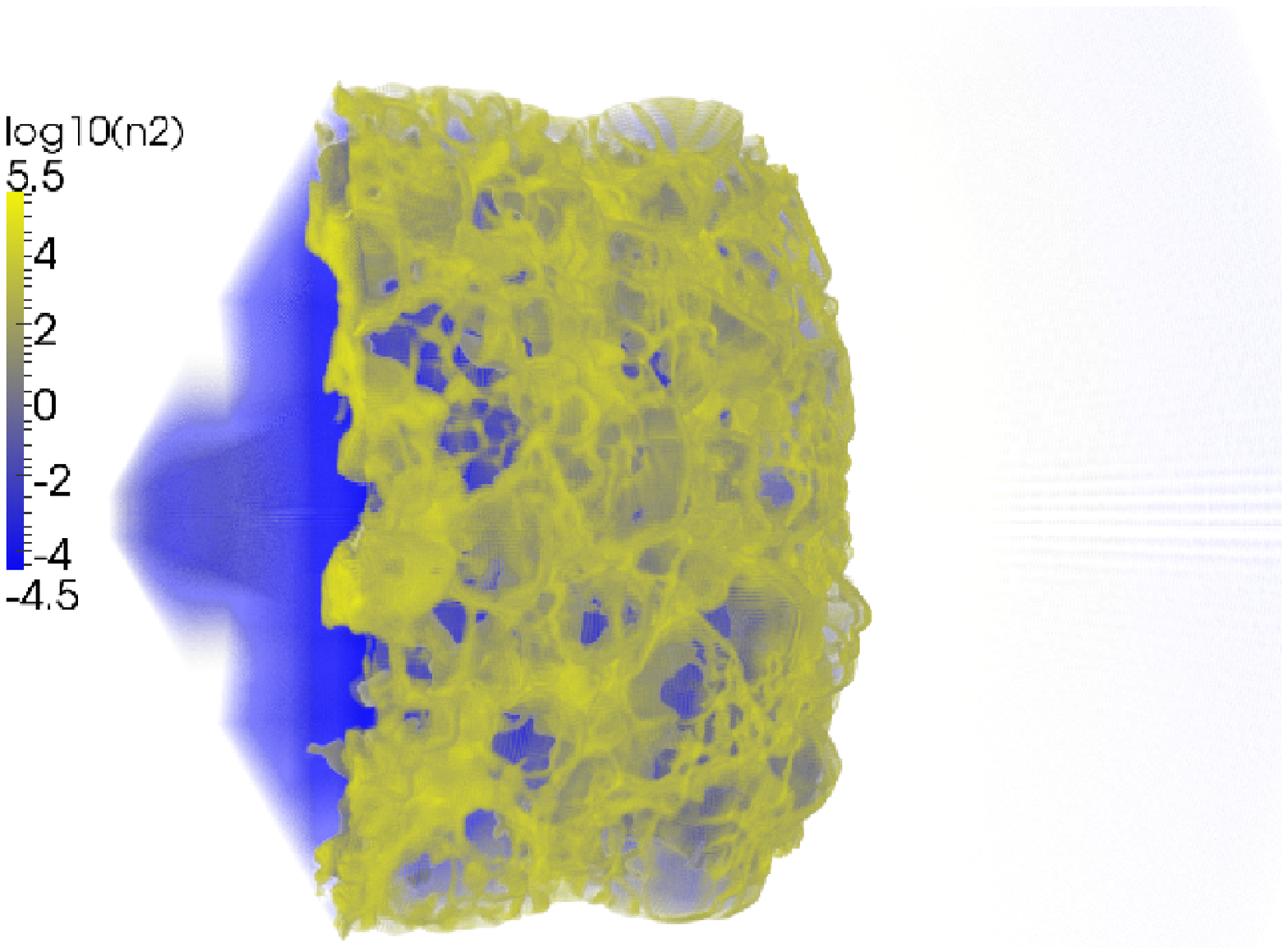}
}
      \caption{Particle density squared [1/cm$^6$] volume rendering of the high resolution WR-RSG interaction (simulation~A4) at the same time as the last panel of Fig.~\ref{fig:WR_RSG_3-D_high}. }
         \label{fig:WR_RSG_high_angle}
   \end{figure*}

   \begin{figure*}
\FIG{
   \centering
   \includegraphics[width=0.95\textwidth]{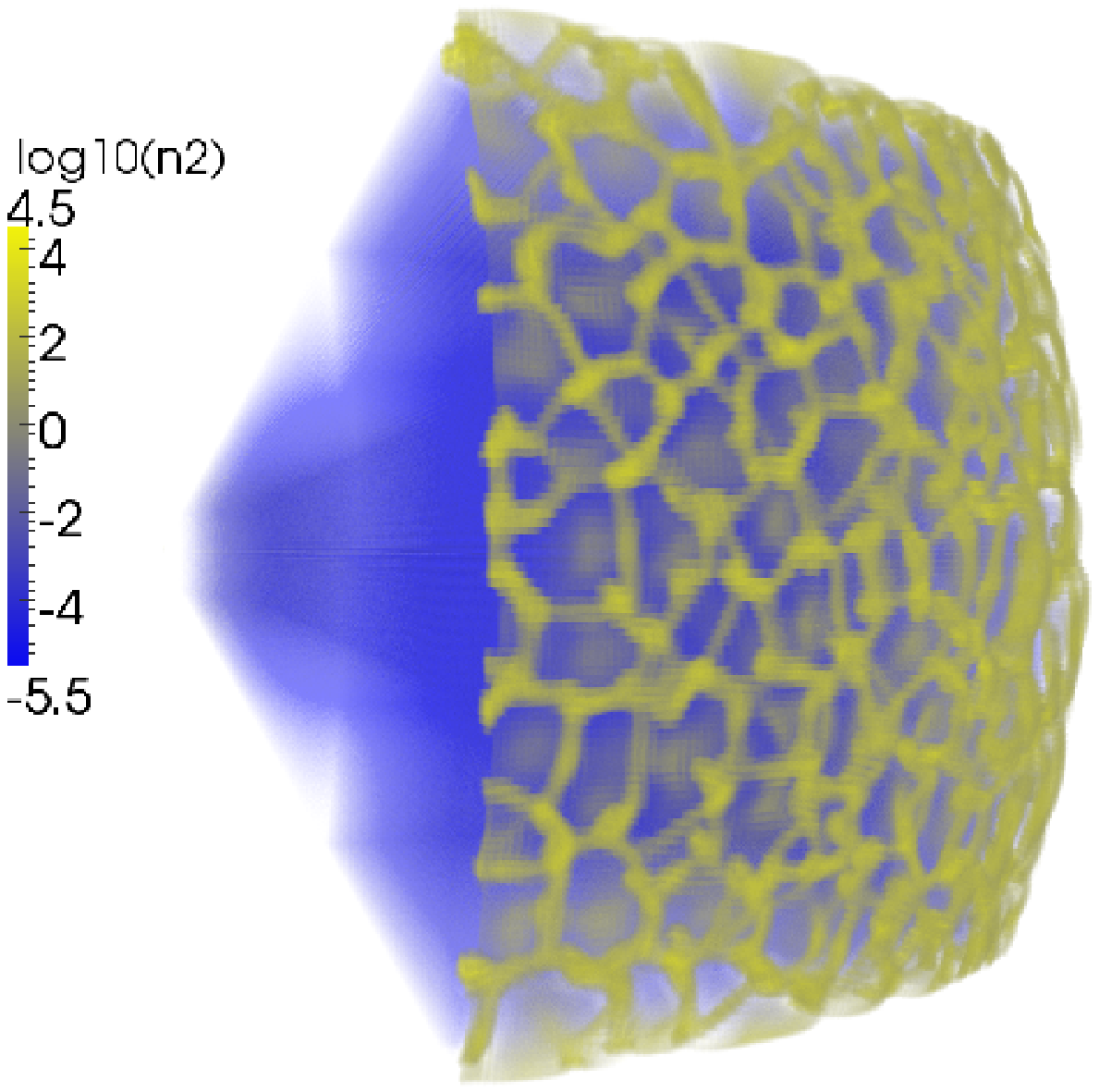}
}
      \caption{Particle density squared [1/cm$^6$] volume rendering of the low resolution WR-LBV interaction (simulation~B3) at the same time as the last panel of  Fig.~\ref{fig:WR_LBV_3-D_low}. }
         \label{fig:WR_LBV_low_angle}
   \end{figure*}

   \begin{figure*}
\FIG{
   \centering
   \includegraphics[width=0.95\textwidth]{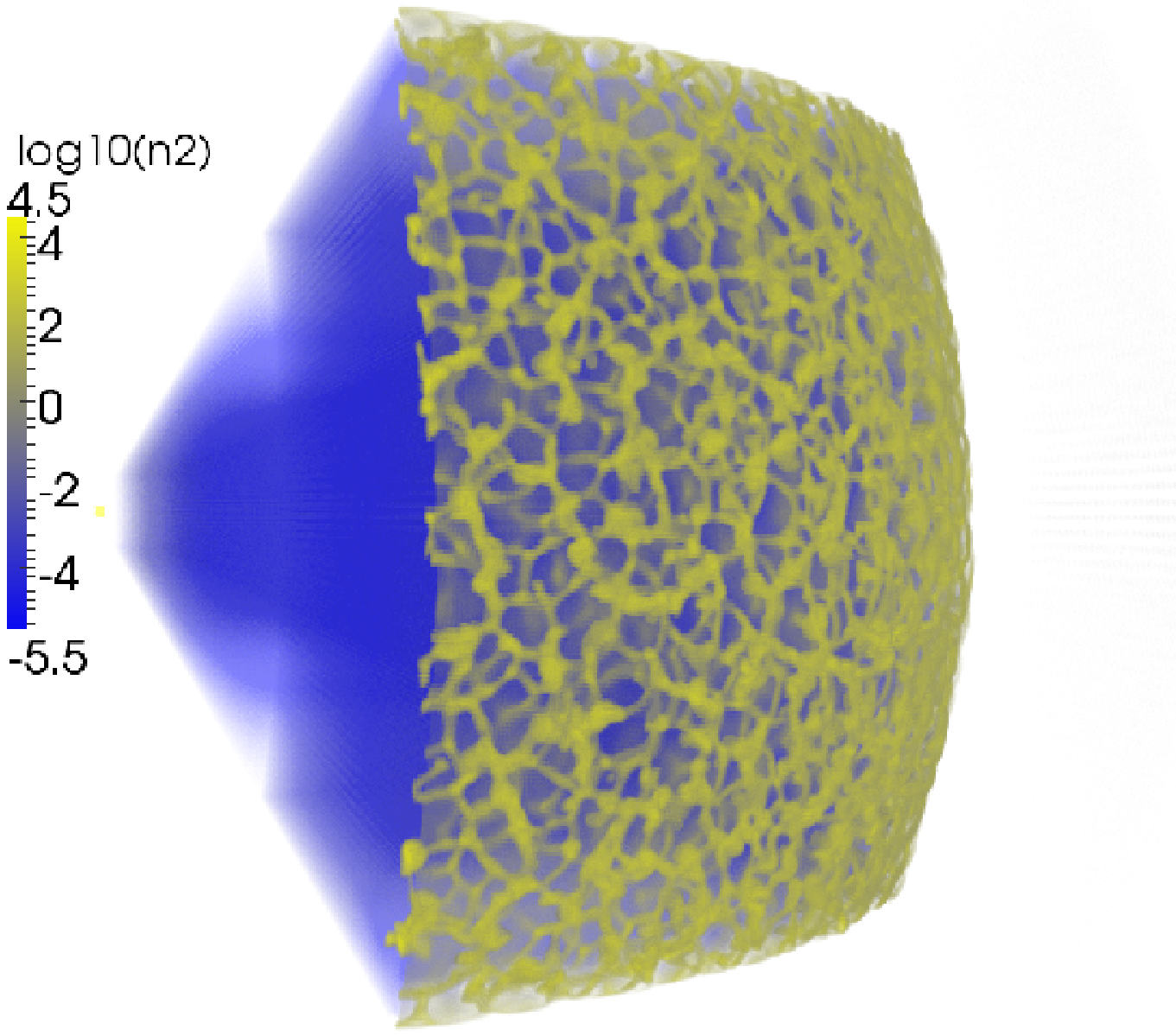}
}
      \caption{Particle density squared [1/cm$^6$] volume rendering of the high resolution WR-LBV interaction (simulation~B4) at the same time as the last panel of  Fig.~\ref{fig:WR_LBV_3-D_high}. }
         \label{fig:WR_LBV_high_angle}
   \end{figure*}

\section{Observable 3-D structure}
\label{sec-observation}
Using our 3-D simulations, we attempt to simulate how the nebulae we produced would appear to an observer. 
Because recombination is the main source of photon production in these emission nebulae and the recombination rate is proportional to the particle density squared, 
we calculate the square of the ion density ($n_h^2\,=\,\rho^2/m_{\rm h}^2$) as a measure of emissivity  \citep[see][]{Garcia-Seguraetal:1999} 
and use it as a basis of a volume rendering of the results shown in the right panels of Figs~\ref{fig:WR_RSG_3-D_low}, \ref{fig:WR_RSG_3-D_high}, \ref{fig:WR_LBV_3-D_low} and \ref{fig:WR_LBV_3-D_high}. 
The result is shown in Figs.~\ref{fig:WR_RSG_low_angle}-\ref{fig:WR_LBV_high_angle}. 
From these figures it becomes clear that the nebula will not appear as a smooth shell. 
Rather, it presents itself as a lattice work of high density filaments with empty space in between. 
Of course, the filaments are connected through a shell, as can be seen in the slices through the 3-D data stack (Figs.~\ref{fig:WR_RSG_3-D_low}-\ref{fig:WR_LBV_3-D_high}). 
However, this shell is so thin, that it does not contribute significantly to the emissivity. 
Only the instabilities, which have a high column depth along the line of sight become visible. 

Typical values for $n_h^{2}$ in the shell are between $10^{2}$ and $10^{5}$. 
To get actual numbers for the emissivity this would have to be multiplied with the local value of $\Lambda(T)$ (see Section~\ref{sec-hydroeqns}). 
The temperature in the shell is typically of the order $10^4$, but since the shock is strongly radiative, the radiative cooling causes a significant drop in the gas temperature, 
causing energy gained through the collision to be radiated away on a timescale shorter than the hydrodynamical timescale of the gas. 
As a result, the temperatures found in the results, which are equilibrium temperatures, may underestimate the actual radiative temperature, 
which should be close to the forward shock temperature ($10^5-10^6$\,K). 
Typical cooling curves show that solar metallicity (a reasonable approximation for the material in the shell, which consists of either LBV or RSG wind), 
would give $\Lambda(T)\sim\,10^{-22}-10^{-21}$\,erg\,cm$^3$\,s$^{-1}$ for these temperatures \citep{DalgarnoMacCray:1972,MacDonaldBailey:1981,MellemaLundqvist:2002,Smithetal:2008,Schureetal:2009}. 
For our simulations this would result in shell-emissivities of the order of $10^{-20}-10^{-16}$\,erg\,cm$^{-3}$\,s$^{-1}$. 
These estimates give an approximation for emissivity in the continuum. 
In order to find line emissivity, one has to use the strength of the emission line, which depends on chemical composition and degree of ionization as well as temperature 
and density. 
Because the shell consists of RSG wind material, which is made up of the stellar envelope, 
the chemical composition can be approximated by the ZAMS composition of the progenitor star. 
(A more accurate estimate can be made based on a stellar evolution model that takes into account chemical mixing due to convection.) 
The degree of ionization is a more complicated issue, as it depends not only on local gas properties, but on  the radiation field of the central star \citep{ToalaArthur:2011}.

Figures \ref{fig:WR_RSG_low_angle}-\ref{fig:WR_LBV_high_angle} show a large difference between the WR-RSG interaction and the WR-LBV interaction. 
The WR-LBV shell appears to have a very regular morphology with little difference between the local structures. 
In Fig.~\ref{fig:WR_RSG_low_angle} instabilities are approximately the same size, with filaments having a width of about 1/100$^{th}$ of a parsec and distances of about 1/10$^th$ of a parsec between knots. 
In Fig.~\ref{fig:WR_LBV_high_angle} the number of filaments is much larger. 
The size of the individual filaments remains very regular, but the distance between knots has become smaller. 
The WR-RSG shell on the other hand, is highly irregular, with large scale instabilities as well as small filaments. 
Here the width of the filaments varies between 1/100$^{th}$ and more than 1/10$^{th}$ of a parsec and they can be more than a parsec in length. 

The effect of a higher resolution, already shown in Section~\ref{sec-results} is again visible, though the effect is quantitative rather than qualitative. 
For the WR-RSG interaction, the low resolution model (simulation A3, Fig.~\ref{fig:WR_RSG_low_angle}) shows less fine structure than the high resolution model 
(simulation A4, Fig.~\ref{fig:WR_RSG_high_angle}), but the morphological trend is clear. 
For the WR-LBV interaction, the general, fairly regular, structure is the same, 
but in the high resolution model (simulation B4, Fig.~\ref{fig:WR_LBV_high_angle}) the individual structures are smaller 
and a secondary network of extremely thin filaments has appeared that could not be fully resolved by the low resolution model (simulation B3, Fig.~\ref{fig:WR_LBV_low_angle}). 
This corresponds to the second peak noted in the Fourier spectrum of simulation~B2.

When comparing these figures to observational results, the most obvious comparison can be made with the ``Crescent nebula'', 
\object{NGC~6888} of which many observations exist in high resolution \citep[e.g.][]{Gruendletal:2000,Mooreetal:2000}. 
This nebula, which is thought to be the result of a WR-RSG interaction shows filaments on a variety of scales (ranging from 1/100$^{th}$ of a parsec to scales close to a parsec). 
Particle densities of clumps in \object{NGC~6888} vary considerably, but \emph{Hubble Space Telescope (HST)} observations show clumps with densities of about $10^3$\,cm$^{-3}$ \citep{Mooreetal:2000}, which coincides with our models, 
which show particle densities ($\rho/m_h$) between $10^2$ and $10^4$. 

\subsection{Observed morphology at a distance.}
\begin{figure*}
\FIG{
\centering
\mbox{
\subfigure
{\includegraphics[width=0.33\textwidth]{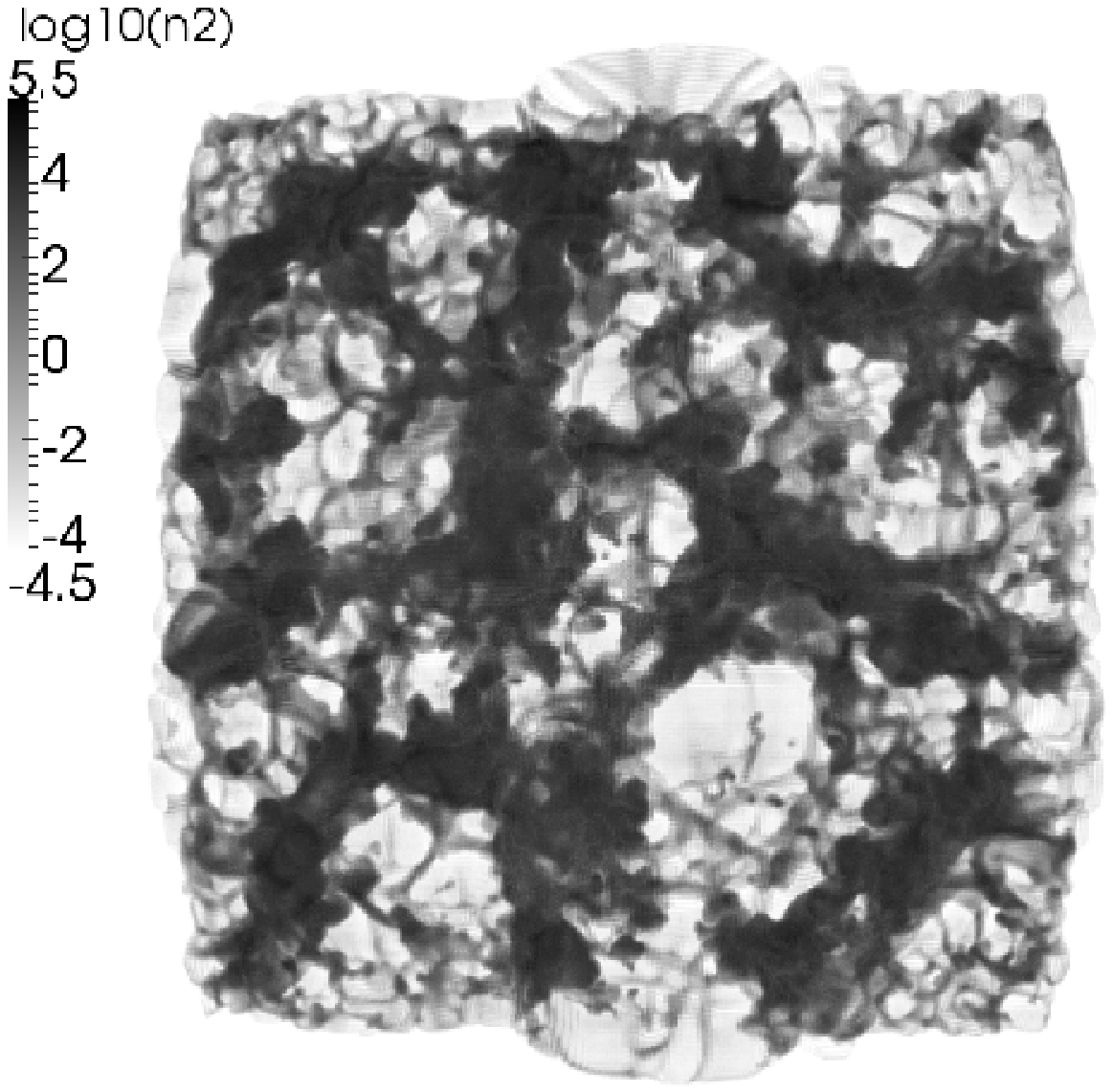}}
\subfigure
{\includegraphics[width=0.33\textwidth]{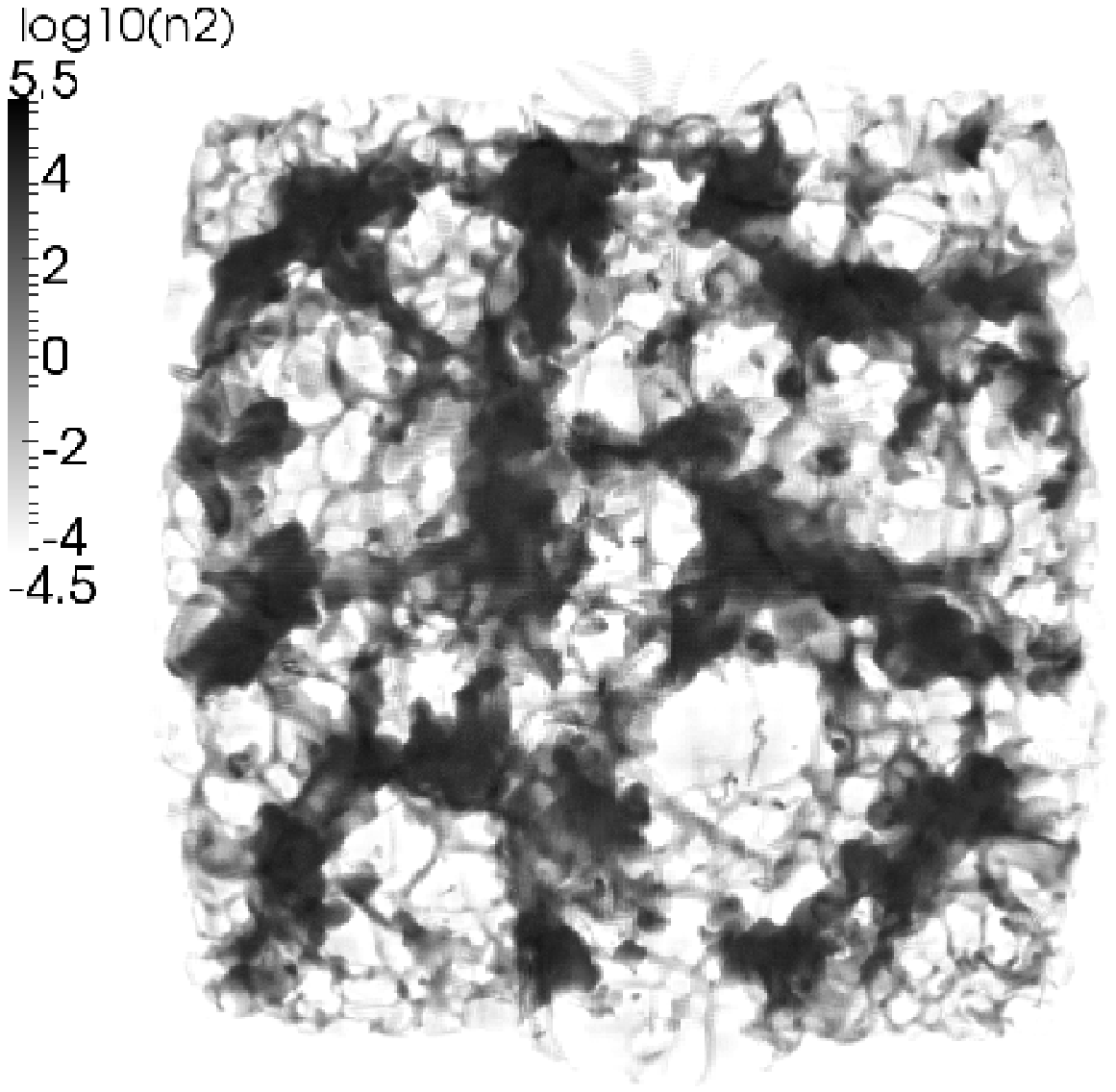}}
\subfigure
{\includegraphics[width=0.33\textwidth]{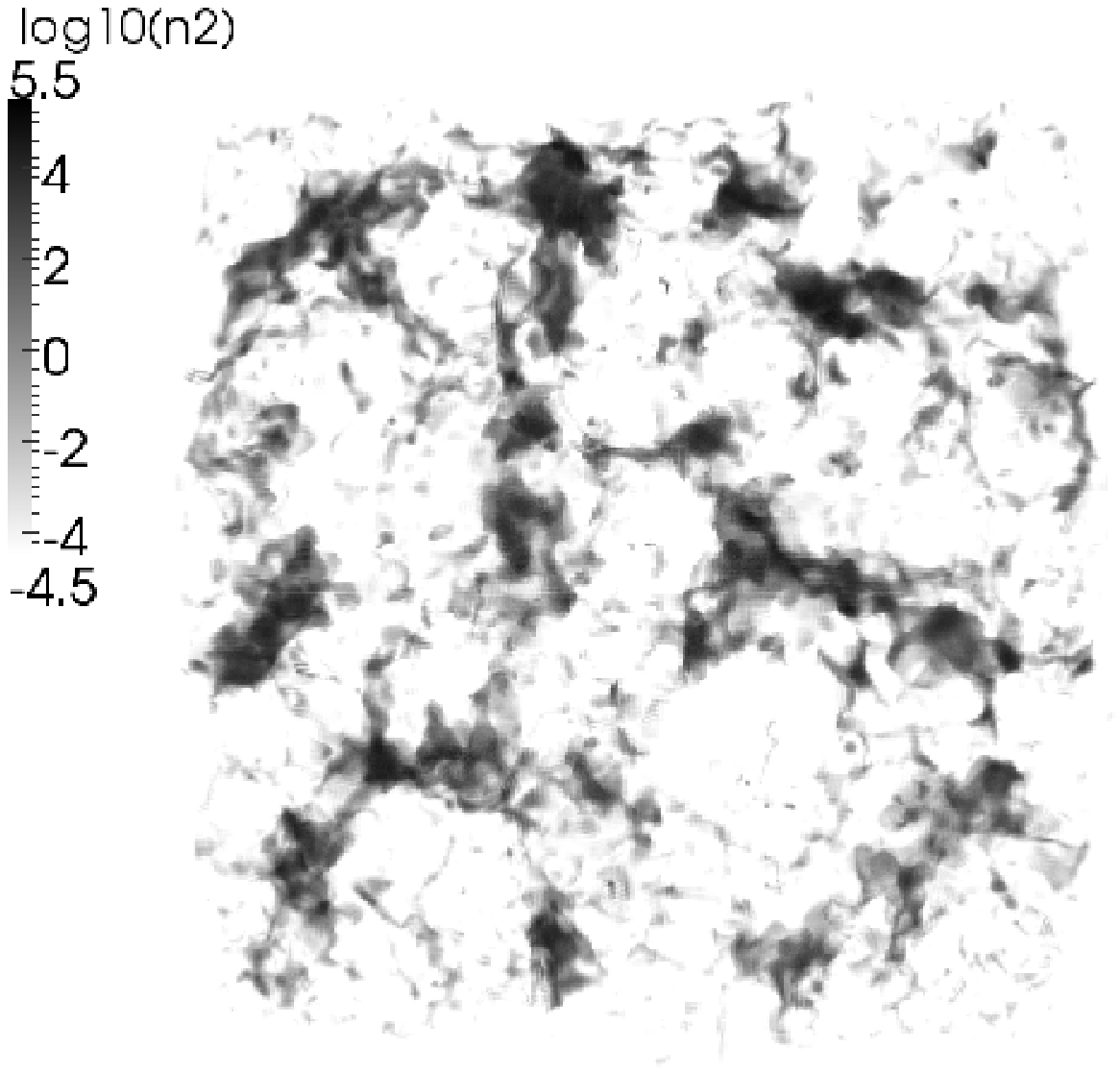}}}
}
\caption{Frontal aspect of the circumstellar shell from Fig.~\ref{fig:WR_RSG_high_angle} (simulation~A4 after 39\,200~years ) 
showing only those parts with $log(n^2)>2$ (left panel) and $log(n^2)>3$ (centre panel) and $log(n^2)>4$ (right panel). 
The larger filaments appear to become thinner and the smaller filaments disappear entirely. The open space between filaments seems to become much larger.}
\label{fig:WR_RSG_visible1}
\end{figure*}

\begin{figure*}
\FIG{
\centering
\mbox{
\subfigure
{\includegraphics[width=0.33\textwidth]{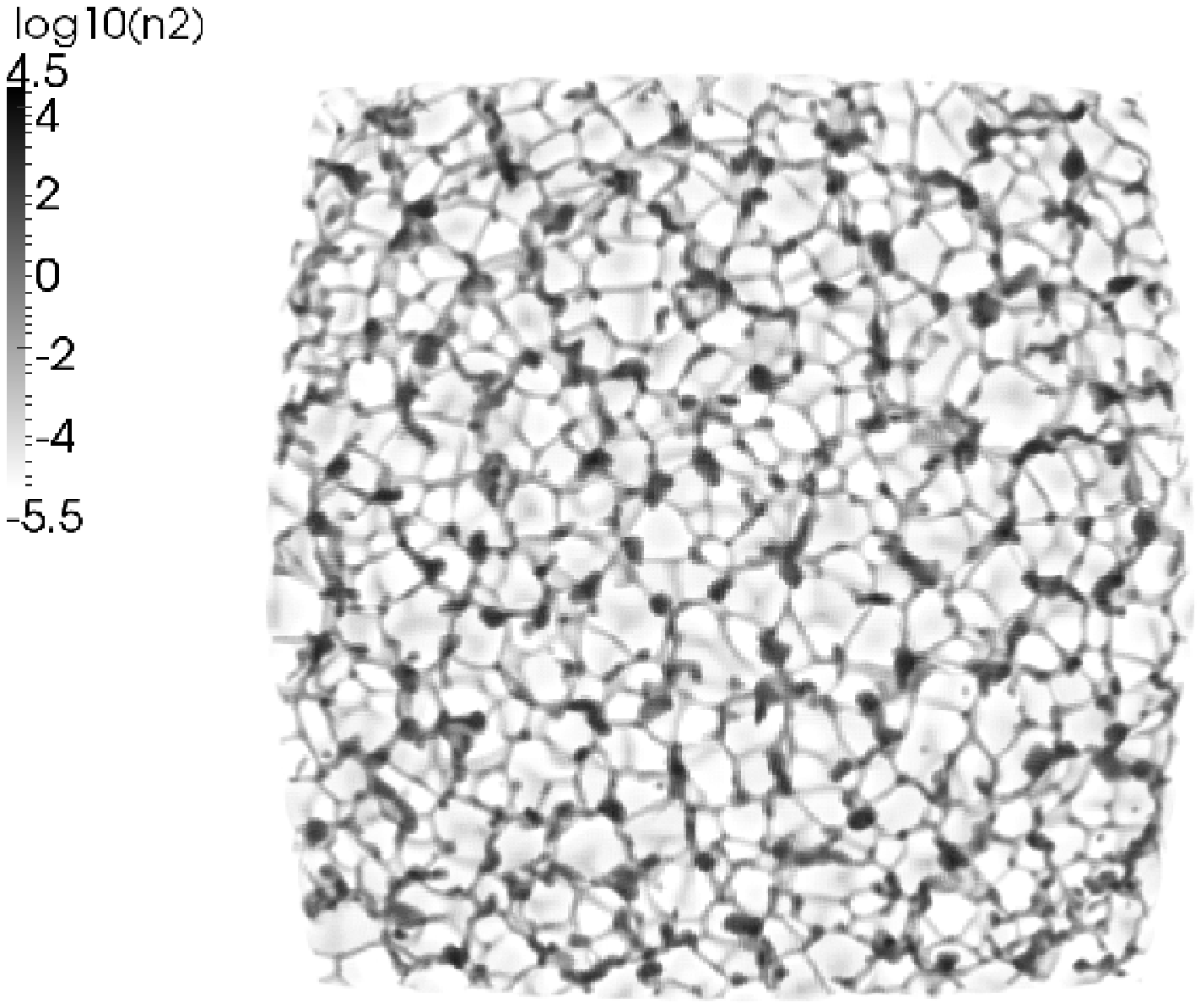}}
\subfigure
{\includegraphics[width=0.33\textwidth]{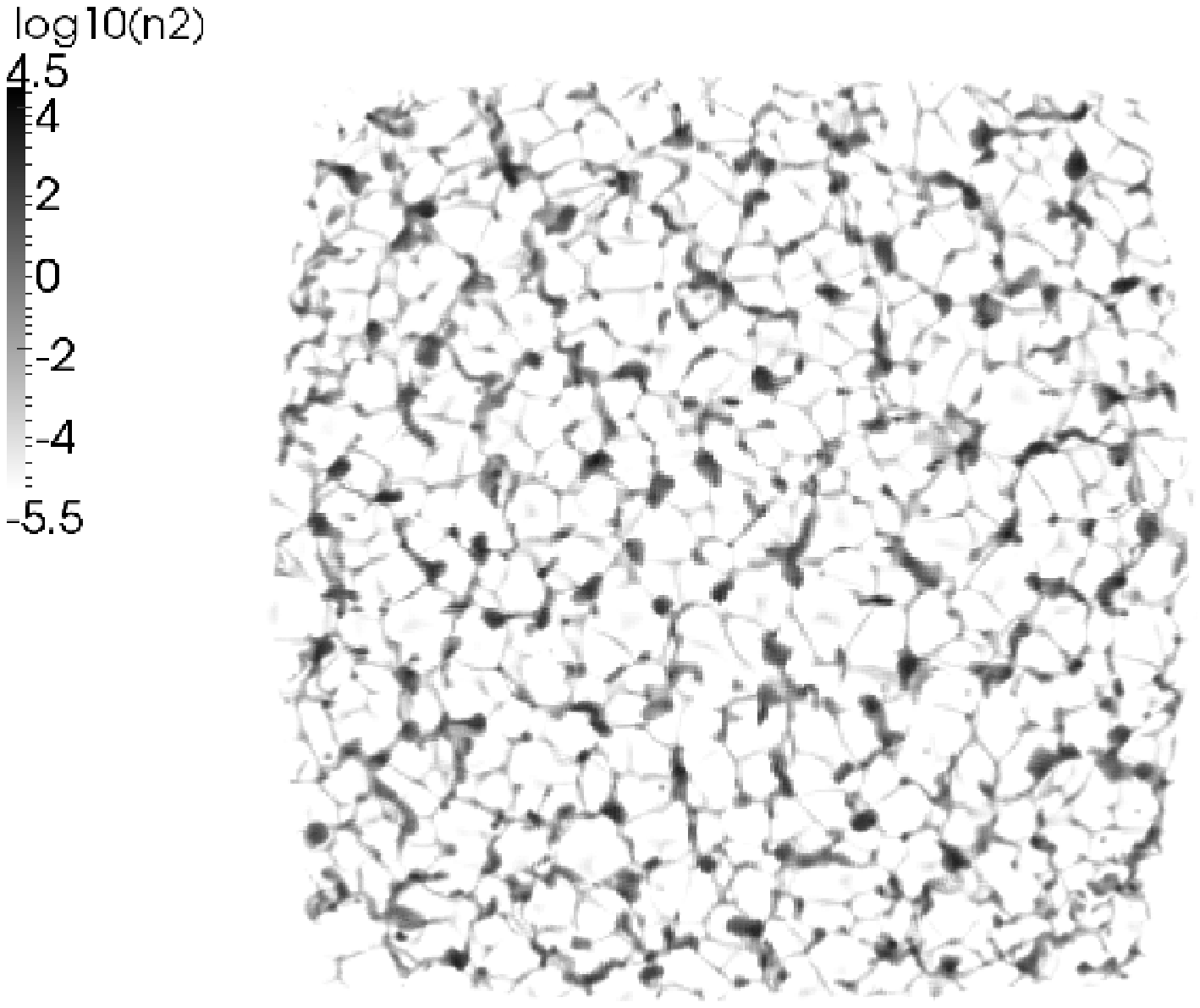}}
\subfigure
{\includegraphics[width=0.33\textwidth]{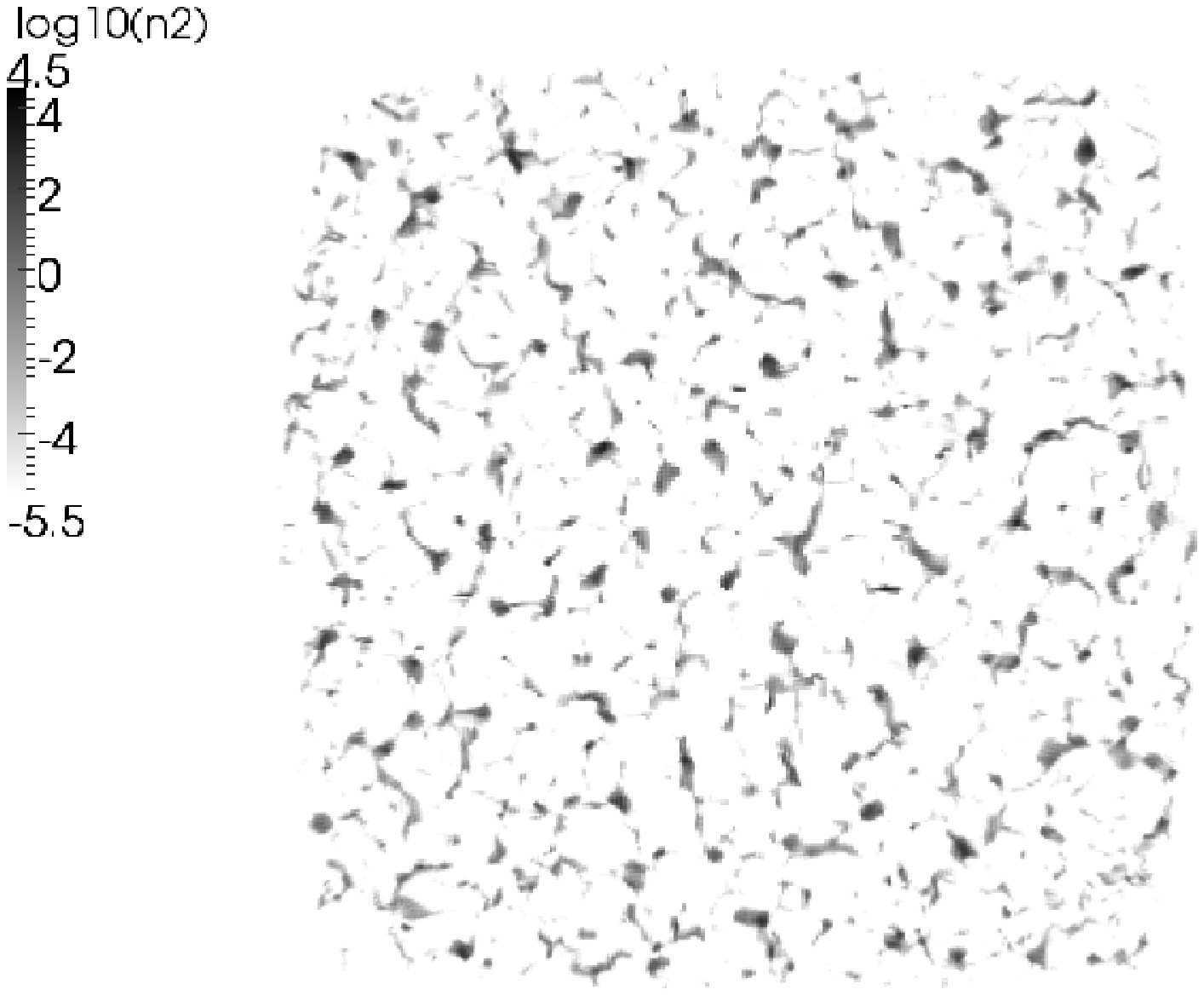}}}
}
 \caption{Frontal aspect of the circumstellar shell from Fig.~\ref{fig:WR_LBV_high_angle} (simulation~B4 after 11\,800~years ) 
showing only those parts with $log(n^2)>1$ (left panel), $log(n^2)>2$ (centre panel) and $log(n^2)>3$ (right panel). 
The qualitative structure does not change much, but the amount of ``empty space'' between filaments increases and the filaments themselves are fading. 
A further reduction in visibility would remove the filaments completely and leave only a handful of knots.}
\label{fig:WR_LBV_visible1}
\end{figure*}

When looking at the 3-D images in Figs.~\ref{fig:WR_RSG_low_angle}-\ref{fig:WR_LBV_high_angle} one should keep in mind the scale of the colour table. 
This varies over ten orders of magnitude for each figure and the shells have a variation of more than three orders of magnitude. 
For these figures, volume rendering has been used to show the entire shell. 
When observing these shells from a distance, only the brightest parts of the shell would be visible, leaving more empty space between the filaments. 
In order to approximate this result, we have made a series of images \ref{fig:WR_RSG_visible1} and \ref{fig:WR_LBV_visible1}, in which we show only part of the gas, 
leaving out those parts where the density becomes too low (and therefore too faint). 
For this we only use the high resolution simulations (A4 and B4). 
 
Figure \ref{fig:WR_RSG_visible1} shows the frontal aspect of the same shell as in Fig.~\ref{fig:WR_RSG_high_angle} with a lower limit of $log(n^2)>2$ (left panel), $log(n^2)>3$ (center panel), and $log(n^2)>4$ (right panel).  
Where originally filaments took up about half of the surface area, the open space now starts to dominate as the larger filaments appear thinner and the smaller filaments disappear completely, 
reducing the visible high density features to about ten percent of the surface area.

For the WR-LBV interaction (simulation~B4), we have to shift the cut-off values, since this nebula is fainter to begin with due to lower densities. 
The results is shown in Fig.~\ref{fig:WR_LBV_visible1} with cut-off values $log(n^2)> 1,\, 2$, and $3$. 
Starting out an order of magnitude fainter than its WR-RSG counterpart, the qualitative structure of this nebula does not change much as it becomes fainter, 
though as in the WR-RSG nebula, the size of the voids between the filaments increases as the lower density parts of the filaments fade away. 
These voids are not actually empty, but represent those parts of the shell, which have no large RT instability. 
Because of this, the cross-section of the high-density region is very short, leading to a low column density. 
Initially the high density filaments covered about ten percent of the surface area. In the final panel, they are reduced to about two percent. 

This analysis does not take into account the limits of spatial resolution. 
For example, the \emph{HST} can resolve structures down to approximately $0.1"$. 
This means that the large filaments of the WR-RSG interaction can be resolved for distances of up to about 2\,Mpc. 
On the other hand, the smallest filaments in the WR-LBV interaction can only be resolved at distances of less than 0.02\,Mpc. 
The \emph{Herschel Space Observatory} infra-red satellite has a spatial resolution that is about a factor 350 less than \emph{HST} 
and can only resolve the small structures at less than 60\,pc. 
Even the larger filaments in the WR-LBV shell would appear would be impossible for \emph{Herschel} to resolve at more than 600\,pc.
Therefore, the shells that result from a WR-LBV wind interaction would appear smooth under most circumstances.

\subsection{Alternative observations} With the launch of satellites like the \emph{Spitzer Space Telescope} and the \emph{Herschell Space Observatory} it has become possible to 
observe the morphology of circumstellar nebulae in the infra-red. 
E.g. Spitzer observations, using the Multiband Photo-Imager (MIPS) \citep{Riekeetal:2004}, show circumstellar shells around \object{Wn8} and \object{Wn9-h} \citep{Mauerhanetal:2010}. 
The main contributor to infra-red emission in circumstellar nebulae is dust. 
Therefore, in order to predict how our models would look in the infra-red we would have to include the presence of dust in our simulations. 
Even without this we can make some predictions. 
\citet{vanMarleetaldust:2011} and \citet{Coxetal:2012} showed that small dust grains, which are by far the most numerous, are tightly bound to the gas through the drag force. 
Therefore, assuming that the dust was initially distributed evenly throughout the RSG wind, 
we can assume that the dust density is directly correlated to the gas density and will show the same morphology. 
This assumption becomes invalid, if significant dust formation takes place in the swept-up shell. 
Should that be the case, the newly formed dust would be concentrated in the high density blobs, because dust formation depends strongly on gas density. 
Dust emission scales with the dust density rather than the density squared, which will make the filamentary structure less prominent. 
Similarly, absorption scales with the density, so any observations that show a circumstellar shell in absorption will show only limited density contrast. 
Still, since the 3-D plots show differences of 5 orders of magnitude or more for $n^2$, 
any observation that scales with $n$ will still show a contrast of several orders of magnitude. 

The large contrasts in column density limit the information on a circumstellar shell that can be gained from the analysis of the spectrum of the central star, 
since the amount of matter observed can change by several orders of magnitude depending on whether the line of sight passes through one of the filaments.

\subsection{The large scale structure of the circumstellar environment}
In the previous analysis we have ignored the larger structure that surrounds the circumstellar nebula. 
Between the nebula and the observer are the unshocked slow wind, the large bubble created by the shocked main sequence wind as well as a shell of interstellar matter 
that has been swept up during the main sequence phase. 
Beyond this shell, is the undisturbed ambient medium. 

The interaction between the main sequence wind and the ambient medium actually reduces the column density and helps make the circumstellar structures visible.
If the star is embedded in a molecular cloud, the column density between the star and the observer will make observations impossible. 
However, during the main sequence phase, the stellar wind sweeps up the surrounding material in a shell, which can break out of the cloud, 
allowing us a direct view of the star and its immediate surroundings. 
This can be seen in the Rosetta Nebula around cluster \object{NGC~2244} \citep[e.g.][]{PhelpsLada:1997,LiSmith:2005}.
The swept-up shell can contain several tens of thousands of solar masses.  
However, this actually increases the visibility, because the swept-up shell contributes less to the column density than if 
the same amount of mass had not been swept up and was still floating in a sphere around the star (see Appendix~\ref{sec-column}). 
The shocked main sequence wind itself does not contribute significantly to the column density. 
The total amount of mass lost during the main sequence is typically $<5\mso$ and spread over a large volume. 
Simulations show main-sequence bubbles of massive stars to have radii of 30-50\,pc 
\citep{GarciaSeguraetal:1996a, GarciaSeguraetal:1996b, Freyeretal:2003, Freyeretal:2006, Dwarkadas:2005, Dwarkadas:2007, vanMarleetal:2005, vanMarleetal:2007, Eldridgeetal:2006, ToalaArthur:2011}.
Assuming an ambient medium with a density of $n_{\rm h}\,=2/{\rm cm}^3$, a main sequence bubble of $R=40$\,pc that contains 5\,$\mso$ of shocked wind material  
and a swept-up shell with a cross-section of 2\,pc, the column density becomes: $9.33\times10^{16}\,n_{\rm h}/{\rm cm}^2$ for the shocked main sequence wind 
and $8.77\times10^{19}\,n_{\rm h}/{\rm cm}^2$ for the swept up shell. 
The column density of the unshocked interstellar medium will, of course, depend on the distance between the observer and the star. 
Inside the shocked main sequence wind, the slow wind (RSG or LBV) also adds to the column density. 
Its contribution is difficult to estimate, because part of it has the $1/r^2$ density profile of the free-streaming wind  
and part is contained in a thin shell against the termination shock of the main sequence wind 
\citep{GarciaSeguraetal:1996a, GarciaSeguraetal:1996b, Freyeretal:2003, Freyeretal:2006, Dwarkadas:2005, Dwarkadas:2007, vanMarleetal:2005, vanMarleetal:2007, ToalaArthur:2011}.

The situation becomes even more complicated when we consider that massive stars typically occur in clusters, 
which increases the amount of matter between the nebula and the observer. 
Still, observations of such nebulae as \object{NGC~6888} already provide as with a detailed view of the structure of the circumstellar shell.  
Infra-red imaging, which can penetrate deeper will improve on this even further.

\section{Discussion}
\label{sec-discussion}
\subsection{Morphology of circumstellar nebulae}
From the results shown in Section~\ref{sec-results}, it is clear that the morphology of a swept-up, circumstellar shell varies greatly depending on  the wind parameters. 
The extremely high density of the slow-moving RSG wind provides far more resistance to the expanding shell than the faster, less dense, LBV wind. 
As a result the shell between the WR wind and the RSG wind is far more unstable than for the WR-LBV interaction. 
A similar effect was noted by \citet{GarciaSeguraetal:1996a,GarciaSeguraetal:1996b}, by \citet{vanMarleetal:2005,vanMarleetal:2007} and by \citet{ToalaArthur:2011}. 
Comparison with these papers shows that our WR-RSG simulations show similar instabilities to previous work, but the RT fingers found in our simulations tend to be broader. 
This can be understood, when we compare the total expansion speed of the circumstellar shell, 
which for our simulations is considerably slower than what was found in earlier papers (See Section~\ref{sec-shellspeed}). 
This lower expansion speed gives the  RT fingers more time to expand sideways. 
Also, because we keep all the gas at a minimum temperature of 10\,000\,K, the thermal pressure in the high density areas remains high. 
\citet{GarciaSeguraetal:1996b} allowed the gas to cool further. 
Therefore, the high thermal pressure in the shocked WR wind would compress the RT fingers more than in our simulations. 
\citet{ToalaArthur:2011} used radiative transfer to determine the degree of ionization (and therefore the temperature) of the gas. 
They find that in the later stages of the WR-RSG interaction the temperature in the high density clumps tends to be at a few 1\,000~K and the RT fingers in their models are more stretched out than ours. 
Our WR-LBV results also resemble previous simulations, but seem less unstable, which is probably due to the lower massloss rates in our model. 
For example, \citet{ToalaArthur:2011} use models for the 60$\mso$ star that have a massloss rate reaching $10^{-3}\,\msoy$. 
The one large instability we find in the high-resolution 2-D model (simulation B2, Fig.~\ref{fig:WR_LBV_2-D_high}) 
is probably a rare event, since it is not found in the 3-D results. 
Also, they may well become less prominent over time as shown in Fig.~\ref{fig:WR_LBV_2-D_extrahigh}.

\subsection{Influence of resolution}
From our simulations, it is clear that a high resolution is very desirable as it can significantly change the result. 
This is especially clear in the 2-D WR-LBV simulations (B1-B2), which show that the low resolution models do not fully resolve the instabilities, 
which inhibits their growth. 
Also, the growth rate of the thin-shell instabilities depends on the thickness of the shell, as discussed in Section~\ref{sec-instabilities}. 
Because the shell in the WR-LBV interaction moves fast, this can leave the instabilities with insufficient time to develop, 
if the resolution is too low to fully resolve the shell. 
A high resolution also allows a higher compression, which in turn increases the density. 
Since RT instabilities depend on density contrast, this increases the growth rate. 
The WR-RSG simulation with its slow moving shell is less vulnerable, 
but the high resolution models show more structure in the instabilities than is found in low resolution. 
The necessity to use high-resolution grids is an important consideration when one has to decide between 2-D and 3-D simulations, 
as high resolution 3-D models will quickly become computationally expensive. 
Our models of the WR-LBV wind interaction show that the shape and size of the individual instabilities does not change significantly between simulation~B2 and 
the model with twice the resolution, so the resolution of B2 is sufficient.

\subsection{2-D vs. 3-D}
The main difference between 2-D and 3-D models seems to lie in the speed at which the instabilities develop. 
That this happens faster for 3-D models can be easily understood. 
If matter moves along the plane of the shell (as is the case for Vishniac instabilities) in 2-D it can only do so along the latitudinal axis, 
since the shell is represented as an essentially 1-D feature. 
In a 3-D model, it can move along a planar shell structure, giving it an extra degree of freedom (the longitudinal axis).
As a result, the local density will vary with the angular velocity squared, rather than as a linear function. 
As a result, at any given time, the instabilities will be more pronounced for a 3-D model than for its 2-D equivalent, as 
was shown by the Fourier analysis. 

The transition from 2-D to 3-D  also influences the shape of the instabilities.  
For example, \citet{Youngetal:2001} showed that RT instabilities in 3-D show more structure than their 2-D counterparts. 
However, the differences are relatively small compared to the total instabilities and will likely prove impossible to observe. 
The behaviour of these instabilities is clearly different from the convection flows described by \citet{MeakinArnett:2007,Arnettetal:2009} and \citet{ArnettMeakin:2011}, 
which showed a marked difference between the 2-D and 3-D simulations.

It is necessary to use 3-D models if one wishes to compare the morphology of the simulated shells directly with observations as shown in  Section~\ref{sec-observation}.
Of course, it is possible to rotate a 2-D result around the polar axis to create a pseudo-3-D grid, but the result would be radically different. 
This technique was used to create projections by \citet{Garcia-Seguraetal:1999} and \citet{Chitaetal:2008} and yields valuable results, 
but instabilities in the 2-D result create rings around the axis when treated in this fashion as can be seen in these papers. 
Only by actually making a 3-D model can we demonstrate what the instabilities would look like.

\subsection{Expansion velocity}
\label{sec-shellspeed}
It is possible to obtain the expansion speed of a circumstellar nebula by measuring the blue-shift of the emission and/or absorption lines created by the swept-up shell. 
Because the expansion speed is a direct result of the wind parameters, 
it can be used as a quantitative means to check the simulation results against observational data.

The expansion speed of the shell can be approximated analytically for a purely adiabatic interaction. 
This predicts an expansion velocity of 
\begin{eqnarray}
V_{\rm s}~=~\biggl(\frac{\dot{m} v^2 {V}}{3\dot{M}}\biggr)^{1/3},
\end{eqnarray}
with $V_{\rm s}$ the bulk motion of the shell, $v$ and $\dot{m}$ the velocity and massloss rate for the fast wind and $V$ and $\dot{M}$ 
the velocity and massloss rate for the slow wind \citep[Eq.\,12.23 from ][]{Kwok:2000}. 
This would give us an expansion speed of 110\,$\kms$ for the WR-RSG interaction and 175\,$\kms$ for the WR-LBV interaction. 
In our simulations we find 85\,$\kms$ and 335\,$\kms$ respectively by measuring the distance travelled by the shell during the simulation.  
(N.B. In the case of the WR-RSG interaction the expansion speed actually varies between 80 and 89\,$\kms$ due to the instabilities distorting the shock. 
The speed of the WR-LBV interaction, which has a smooth forward shock, can be determined with an accuracy of about 1\,$\kms$.) 

For the WR-RSG interactions, the analytical prediction is too high. 
This is partly due to the extensive instabilities in the WR-RSG simulations, which use up part of the available energy. 
Furthermore, due to the high densities involved, the WR-RSG interaction has strong radiative cooling, which allows energy to leak out of the system. 
The analytical value for the speed of the WR-LBV interaction is clearly impossible, since it is actually less than the wind velocity of the LBV wind, 
so according to this prediction there would not be a shell. 
The explanation lies in the assumptions on which the analytical model is based. 
Specifically, \citet{Kwok:2000} assumed that the mechanical energy of the slow wind is much less than for the fast wind. 
This is correct for the AGB to post-AGB transitions that form planetary nebulae as well as for the WR-RSG interaction. 
However, for the WR-LBV interaction, both winds have similar mechanical energy.

The expansion velocity we find for the WR-RSG interaction is significantly lower than the values found by e.g. \citet{GarciaSeguraetal:1996b,vanMarleetal:2005,Freyeretal:2006} for a similar situation. 
These models used wind parameters based on the massloss rates from \citet{deJageretal:1988}, 
which tend to overestimate the massloss rates for hot stars as they ignore the effect of clumping in the stellar wind. 
Our models use a lower WR massloss rate, based on the work of \citet{Bouretetal:2005,Vinkdekoter:2005,Mokiemetal:2007}, which take the effect of clumping into account. 
Observations of WR wind nebulae typically give low ($\sim 50-100\,\kms$) expansion velocities for the WR-RSG interaction \citep[E.g.][]{Chuetal:1983,Goudisetal:1988,Smithetal:1988,DysonGhanbari:1989,Gruendletal:2000}.
This seems to confirm the validity of our results.

\section{Conclusions}
\label{sec-conclusions}
In the past, 2-D hydrodynamical simulations have been successful in reproducing  the circumstellar shells of massive stars. 
We have now extended these models to 3-D. 
We find that, although the 2-D models resemble the 2-D cuts made through the 3-D results, 
the 2-D models are insufficient to fully model the structure of circumstellar nebulae. 
Specifically, local density fluctuations in the shell that result from hydrodynamical instabilities appear in 2-D models as individual clumps. 
In reality, the instabilities form a lattice of filaments that can only be reproduced in a 3-D model. 
Even the slices tend to show more pronounced instabilities than the 2-D models as shown in their respective power spectra, 
though this may be nearly invisible.

The expansion speed of the circumstellar shell in the WR-RSG interaction is markedly slower in our simulations 
than found by \citet{GarciaSeguraetal:1996b,vanMarleetal:2005,Freyeretal:2006}, 
which can be accounted for by the lower RSG wind velocity (10\,$\kms$ vs. $15\,\kms$) 
and lower massloss rate during the WR phase ($10^{-5}\,\msoy$ vs. ($10^{-4.5}\,\msoy$). 
Since our expansion velocity ($<100\,\kms$) lies closer to observed nebula expansion velocities this can serve as an additional validation for the lower WR massloss rates. 
The lower expansion speed also has consequences for the shape of the instabilities, which have more time to form and expand, 
leading to broader, more irregular shapes than found in previous models. 

For the future we hope to expand our work to include stellar rotation \citep{Chitaetal:2008,vanMarleetal:2008} 
as well as the interaction between the winds of binary companions \citep[e.g.][]{Pittard:2009,PittardParkin:2010,vanMarleetal:2011b} and massive stars in clusters \citep{vanMarleetal:2012}. 
We also intend to add additional physics, such as the influence of dust \citep{vanMarleetaldust:2011,Coxetal:2012}.

\begin{acknowledgements} 
We thank Lilith Axner and Wim Rijks at SARA Computing and Networking Services (http://www.sara.nl/).
A.J.v.M.\ acknowledges support from NSF grant AST-0507581, from the
FWO, grant G.0277.08 and K.U.Leuven GOA/09/009 and from the DEISA Consortium (www.deisa.eu), co-funded through the EU FP7 project RI-222919, for support within the DEISA Extreme Computing Initiative. 
Simulations were done at the Flemish High Performance Computer Centre, VIC3 at K.U. Leuven and at the CINECA-SP6 in Bologna, Italy.
A.J.v.M. thanks dr.~Z.~Meliani for his help with setting up the MPI-AMRVAC code. 
Images were created with the Paraview (http://www.paraview.org/), Visit (https://wci.llnl.gov/codes/visit/) and Tecplot$^{\rm{\small TM}}$ (http://www.tecplot.com/) visualization software.
FFT was computed with IDL 7.0 (http://www.exelisvis.com/language/en-US/ProductsServices/IDL.aspx). 
This research has made use of NASA's Astrophysics Data System. 
We thank the anonymous referee for the many helpful comments and suggestions that allowed us to improve our paper.

\end{acknowledgements}

\bibliographystyle{aa}
\bibliography{18957}

\IfFileExists{18957.bbl}{}
 {\typeout{}
  \typeout{******************************************}
  \typeout{** Please run "bibtex \jobname" to obtain}
  \typeout{** the bibliography and then re-run LaTeX}
  \typeout{** twice to fix the references!}
  \typeout{******************************************}
  \typeout{}
 }

\begin{appendix}

\section{Fourier analysis of instabilities}
\label{sec-fft}
We use Fourier analysis \citep[][p.283-287]{Russ:1995} to quantify the instabilities in circumstellar shells (See Section~\ref{sec-results}). 
This was done using the fast-Fourier transform (FFT) routine provided by {\tt IDL 7.0}. 
Calculating the Fourier transform as shown in Figs.~\ref{fig:WR_RSG_fft}-\ref{fig:WR_LBV_fft} requires the following steps. 
First of all, we map the data onto a uniform grid with the same resolution as the highest level of the adaptive mesh. 
This is accomplished using the same interpolation algorithm as used by {\tt MPI-AMRVAC} for the adaptive mesh during the simulation. 
We calculate the particle density according to $n=\rho/m_h$. 
The particle density is a more practical quantity than the mass density, because the values are close to one. 
We then isolate the part of the grid containing the circumstellar shell.  
In this region we calculate a 1-D FFT along each latitudinal gridline and take the power spectrum, which is given by:
\begin{equation}
P(\nu) = |S(\nu)|^2
\end{equation}
with $P(\nu)$ the power spectrum and $S(\nu)$ the result of the FFT. 
We then take the average of the 1-D results and normalize it to the height of the zero-order maximum. 
The result is then plotted as shown in Figs.~\ref{fig:WR_RSG_fft}-\ref{fig:WR_LBV_fft}.

\section{Column density of a swept-up shell}
\label{sec-column}
As the main sequence wind of a star sweeps up the ambient medium, the effective column density changes, 
 because the matter is concentrated in a shell. 
As the shell expands, the surface over which the matter is spread out increases, reducing the amount of mass that a ray encounters 
as it passes through the shell. 
In the case of undisturbed matter the column density is simply $ \rho_c~=~\rho_{\rm ISM}\,R$, 
with $\rho_c$ the column density, $\rho_{\rm ISM}$ the ambient medium density and $R$ the distance to the star. 
If the same amount of mass has been swept up in a shell with a thickness of $\Delta R$ that extends from $R-\Delta R$ to $R$, the column density becomes
\begin{eqnarray}
 \rho_c~&=&~\frac{\frac{4}{3}\pi \rho_{\rm ISM} R^3}{4\pi (R-\Delta R)^2 \Delta R}\,\Delta R, \\
         &\approx&~\frac{1}{3} \rho_{\rm ISM} \frac{R^2}{R-2\Delta R}, 
\end{eqnarray} 
for $\Delta R\,\ll\,R$.
Therefore, the column density in the swept up shell is less than that of the same medium in undisturbed condition if
\begin{eqnarray}
 \rho_{\rm ISM}\,R~&>&\frac{1}{3} \rho_{\rm ISM} \frac{R^2}{R-2\Delta R} ,\\
               R~&>&3\Delta R.
\end{eqnarray} 
For the bubbles around massive stars this is the case, so we can conclude that the main sequence wind reduces the column density by sweeping up the ambient medium.

\end{appendix}

\end{document}